\def\pp{\noindent\parshape 2 0truecm 17truecm 2truecm 15truecm}
\def\rf#1;#2;#3;#4 {\par\pp#1, #2, #3, #4. \par}

\def\pr{\ref@jnl{Phys.Rev}}     

\def\etal{{\frenchspacing\it et al. }}

\newcommand{\href}[2]{{\bf #2} (\texttt{#1})}

\def\dex{\rm dex}
\def\beq#1{\begin{equation}\label{#1}}
\def\eeq{\end{equation}}
\def\beqa#1{\begin{eqnarray}\label{#1}}
\def\eeqa{\end{eqnarray}}

\def\tento#1{\times 10^{#1}}

\def\K{{\rm \ K}}
\def\s{{\rm \ s}}
\def\sr{{\rm \ sr}}

\def\erg{{\rm \ erg}}
\def\cm{{\rm \ cm}}
\def\eV{{\rm \ eV}}
\def\Mpc{{\rm \ Mpc}}

\def\nd#1{n_{ \rm #1}}
\def\k#1 {k_{{\rm #1}}}

\def\HH{H$_2$ }
\def\HHp{H$_2^+$ }
\def\Hp{H$^+$ }
\def\Hm{H$^-$ }
\def\Hep{He$^+$ }
\def\Hepp{He$^{++}$ }

\def\mHH{H_2}
\def\mH2p{H_2^+}
\def\mHp{H^+}
\def\mHm{H^-}
\def\mHep{He^+}
\def\mHepp{He^{++}}

\documentstyle[epsfig, 11pt,aj_pt]{article}

\begin{document}
\bibliographystyle{plain}
\title{\bf \sc Modeling primordial Gas  in \\ Numerical Cosmology}
\author{Tom Abel, Peter Anninos,  Yu Zhang \& Michael L. Norman\\ 
Laboratory for Computational Astrophysics at the National Center for
Supercomputing Applications \\ 
        {\small abel@astro.uiuc.edu\\
        panninos@ncsa.uiuc.edu \\
        yzhang@ncsa.uiuc.edu \\
        norman@ncsa.uiuc.edu}}
\authoraddr{5261 Beckman Institute, Urbana, Ill, 61801}
\date{\today}
\thispagestyle{empty}

\begin{abstract}
  We have reviewed the  chemistry and cooling behaviour of low-density
  ($n<10^4\cm^{-3}$) primordial gas  and devised a cooling model  wich
  involves 19 collisional and 9  radiative processes and is applicable
  for  temperatures in the range (1K$  < T <  10^8$K).  We derived new
  fits of rate   coefficients   for the photo-attachment of    neutral
  hydrogen, the formation  of   molecular  hydrogen via \Hm,    charge
  exchange between  \HH  and \Hp,  electron   detachment of  H$^-$  by
  neutral hydrogen, dissociative   recombination of H$_2^+$ with  slow
  electrons, photodissociation  of  H$_2^+$,  and photodissociation of
  H$_2$.  Further it  was found that  the molecular hydrogen  produced
  through the  gas-phase processes, \HHp + H  $\rightarrow$ \HH + \Hp,
  and \Hm +  H $\rightarrow$ \HH + e$^-$,   is likely to be  converted
  into  its para configuration   on   a faster  time  scale than   the
  formation time scale.  We    have tested the model   extensively and
  shown it to agree well with  former studies.  We further studied the
  chemical kinetics in great detail and  devised a minimal model which
  is substantially simpler than the full reaction network but predicts
  correct abundances.  This minimal model  shows convincingly that  12
  collisional processes  are sufficient to model the  H, He, \Hp, \Hm,
  \Hep, \Hepp,  and \HH abundances  in low  density primordial gas for
  applications with no radiation fields.
\end{abstract}
\keywords{atomic processes ---  molecular  processes  --- plasmas  ---
 IGM: molecules  ---    radiation mechanisms: thermal    --- radiative
 transfer }
\thispagestyle{empty}

\section{Introduction}

Since Szalay (1967) realized the importance of the gas phase reactions
for  \HH formation in primordial gas
\begin{eqnarray}\label{equ:H2paths}
\rm
H \  \ \ \  + \ \ e^-  \  & \rightarrow & \ \ {\rm \mHm} \ \  +  \ \ h\nu,  \\
\rm \mHm  \ \ + \ \ H \ \ & \rightarrow & \ \ \rm \mHH  \ \ \  +  \ \  e^-, 
\end{eqnarray}
and
\begin{eqnarray}
\rm
\mHp \ \ + \ \ H  \ \ & \rightarrow & \ \ {\rm \mH2p} \ \  +  \ \ h\nu,  \\
\rm \mH2p  \ \ + \ \ H \ \ & \rightarrow & \ \ \rm \mHH   \ \  +  \ \ \mHp,
\end{eqnarray}
and  Peebles and Dicke  (1968)  formulated their  theory  for globular
cluster  formation based on   \HH   cooling of collapsing   primordial
density fluctuations, many models have been put forward employing this
effect to explain the  origin of  structures on a  vast range  of mass
scales.  E.g.,   for the cosmic  string  model of structure formation,
molecular  hydrogen is believed to  trigger  the fragmentation of late
cosmic string wakes.   This is because,  for the weak accretion shocks
encountered during the formation of  late cosmic string wakes, thermal
instabilities due to hydrogen  line, or bremsstrahlung cooling are not
accessible   (Rees  1986).  Kashlinsky  and   Rees  (1983)   discussed
fragmentation of primordial gas that cools via \HH.  Couchman and Rees
(1986) envisaged star cluster sized objects collapsing via \HH cooling
at  high redshift which then  reionize the intergalactic medium.  More
recently,  a  model  for  the  origin  of halo  globular  clusters and
spheroid  stars    based on a cooling  instability    triggered by \HH
formation     has been  put  forward  by   Vietri   and Pesce (1995).  
Furthermore, in   studies of   primordial star   formation,  molecular
hydrogen  cooling plays a central role  (see Stahler 1986 for review). 
All   these applications   are  of  fundamental   importance  for  our
understanding of  the  origin of  structure  in the  universe  and its
subsequent evolution.

All  of the   above  problems are inherently  multi-dimensional  and a
method that  computes  the hydrodynamics along  with  the chemistry is
desirable.  In order  to form \HH   efficiently through the  above gas
phase reactions, electrons   and   protons  have to be   abundant   at
relatively low temperatures ($T<10^4\K$) allowing the formed molecules
to  survive.  This situation  arises  naturally  in shock  heated gas,
where the  gas  recombines slower than   it can cool, and an  enhanced
ionized fraction (over the equilibrium  value) is reached, despite the
low  temperatures.  This has been convincingly  shown by many previous
investigations   (Hollenbach and McKee   1979, MacLow  and Shull 1986,
Shapiro  and Kang 1987,   Anninos and  Norman 1996).   Furthermore, at
recombination the universe expands so fast that recombination will not
be  complete and the gas  is  left with  a  residual abundance of free
electrons (see Peebles 1993).  This  residual ionized fraction can  be
enough to form  a substantial amount  of  \HH in structures  that only
have   very  weak virialization shocks  (see   Couchman and Rees 1986;
Tegmark  \etal 1996).  However,    the  chemistry has  to  be   solved
self-consistently with  the  structure formation   equations, which is
computationally due to the stiff nature of the reaction network.  This
has forced former studies to constrain  themselves to zero dimensions,
using e.g.  steady state  shock approximations.  Only  recently Haiman
\etal presented a   study  of collapsing  small  scale structure  that
incoorporated  the   time  dependent chemistry  in  a  one dimensional
hydrodynamics code.  In  our methodology paper (Anninos,  Zhang, Abel,
\& Norman 1996, hereafter AZAN96)  we discuss a numerical method  that
unifies   the   time    dependent   chemistry  with  multi-dimensional
cosmological hydrodynamics.  With the chemistry model which accurately
predicts abundances, and  the cooling behaviour  of primordial  gas for
temperatures from one  to  $\sim 10^{8}$  Kelvin and all  densities
below   $\sim   10^{4}\cm^{-3}$,  this  introduces   a    powerful  tool in
investigating many aspect  of structure formation   by means of  three
dimensional hydrodynamical simulations.

The paper is organized  in the following way:  In the next  section we
review    some  general   properties  of     primordial gas  such   as
nucleosynthesis   constraints  and  the  coronal   limit.  In  section
\ref{sec:SelRC}  we  discuss general arguments  on   how to select the
dominant reactions.  We  then give an   elaborate presentation of  the
collisional and radiative processes which  we find to be important and
the  rate coefficients  we have adopted.   Section \ref{sec:Neglected}
gives an overview of  the processes we  find to be negligible for  the
range of applications we are interested in. Then we review all cooling
mechanisms needed for our model and discuss  the molecular cooling and
heating  functions in the   low-density limit.  Finally we  present an
extensive discussion   of   the performance  of   our  model  for   an
application to  a strong  shock wave  arising  from the collapse  of a
single pancake or cosmological  sheet.  The complete chemical model is
summarized in appendix  \ref{app:rates}.   The discussion why we  find
all \HH to  be  in  its  para   configuration  is given  in   appendix
\ref{sec:OtPRatio}.

\section{General Properties of Primordial Gas}

What are the major species determining the  physics in primordial gas?
Here  we briefly  summarize what we   include in our cooling model  and
defer the detailed discussion to the following sections.

We  know  from  nucleosynthesis and  observational   constraints  that
$^7$Li/H $\sim 10^{-10}$, D/H$\sim 10^{-5}$, $^3$He/H $\sim 10^{-5} $,
and 0.236 $\leq$ $^4$He/H $\leq$ 0.254.  Where H, D, He, Li denote the
mass    abundances  of  hydrogen,     deuterium, helium    and lithium
respectively.     Hence D, $^3$He,  and  Li  have, compared to neutral
hydrogen, very low abundances.  For the line cooling of D, $^3$He, and
Li to contribute  significantly to  the overall  cooling of  the  gas,
their    excitation rates would have to    be many orders of magnitude
higher than the corresponding processes  of neutral hydrogen. However,
this is not  the case and those species,  at least in their atomic and
ionic forms, will have  negligible  influence on the hydrodynamics  of
the gas.   The work by Lepp  and Shull (1984) indicates that molecular
hydrogen was  by  far the most  abundant  molecule in  the universe at
redshifts between recombination and before the first stars formed.  It
forms  by  radiative association of   the  negative hydrogen ion  with
neutral hydrogen  and  charge  exchange  between H and   H$_2^+$.  The
minimal temperature  at which rot./vib.  of  molecules can get excited
defines a minimal temperature, and hence a  minimal Jeans mass. In the
case of  molecular hydrogen, this introduces  a minimum temperature of
about $10^2$K (see also Mac Low and Shull 1986, Shapiro and Kang 1987)
which translates to a Jeans mass of
\begin{eqnarray}
M_J & = &\left(\frac{\pi k T}{\mu G} \right)^{\frac{3}{2}} 
                \rho^{-\frac{1}{2}}                             \\
    & = & 2\times 10^5 {\rm M}_\odot 
                \left( \frac{T}{100{\rm K}}\right)^{\frac{3}{2}} 
                \left( n/{\rm cm}^{-3} \right)^{-\frac{1}{2}}, 
\end{eqnarray}
which tells us that in an application as e.g. proto-galaxies, where we
expect number densities of the order $100$cm$^{-3}$  we expect a Jeans
Mass of about $10^4$M$_\odot$. This is a rough estimate of the minimum
mass  scale one   has  to resolve in  numerical   simulations.  If one
includes also HD and LiH which will get  excited only for temperatures
above  $112$K and $21$K,  respectively,  one has  be  able  to resolve
masses nearly a magnitude smaller.  A study by D.   Puy et al.  (1993)
was able to give fairly good estimates on  the primordial HD abundance
after recombination.  They found that HD was  the second most abundant
molecule with   $n_{HD} \sim 10^{-3.5}n_{H_2}$.   We know,  therefore,
that molecular hydrogen will  always be the dominant molecular coolant
for temperatures above  $\sim  100$K.  Hence  choosing to not  include
other molecules   than H$_2$ can  be understood  that we  restrict our
attention  to   masses  greater than    $\sim 10^4$M$_\odot (n/100{\rm
cm}^{-3})$.   Recently the Lithium  Chemistry has been rigorously 
discussed by Stancil \etal 1996, where they found that  much less
primordial LiH is formed than previously expected and suggested that a
significant   fraction  of LiH might  only   be  formed  by three body
reactions.   Since our model  focuses at  number densities below $\sim
10^{4} \cm^{-3}$ and  three--body reactions typically become important
at greater densities  we  conclude that  in the density  regime we are
interested in LiH  will not be important.   The papers mentioned above
did not include  H$_3^+$   which, due to  its  low  abundance  is  not
believed  to be  a  significant  coolant.   From the  calculations  by
Lenzuni,   Chernoff,  and Salpeter (1991) we    know that  in fact the
H$_3^+$ abundance can be of the same order  as the H$_2^+$ abundance,
but  their results also show that  it has a   negligible effect on the
molecular hydrogen abundance.  We   also did not include  He molecular
ions as well as hydrogen ion  clusters (H$_n^+$ with $n\geq4$) because
their  expected  abundances  are even   smaller  than the ones  of the
molecules mentioned above.

\subsection{``Molecular'' and ``Atomic'' Coronal Limit}
\label{sec:coronallimit}  

Atoms and molecules have very complex  intrinsic properties, and their
chemical  behavior   varies  somtimes drastically  with  the  specific
quantum-mechanical state they occupy.   For atoms and ions at moderate
or  low  densities like in  the  solar corona (electron number density
$n_e    \sim   10^8-10^9$cm$^{-3}$),   the  following    features   of
thermodynamical equilibrium do {\em not} hold (Sobelman et al. 1979):

\begin{itemize}
\item{Boltzmann distribution of atoms over excited states.}
\item{Saha distribution of atoms over degrees of ionization.}
\item{Principle of detailed balance.}
\end{itemize}

The velocity distribution of the free electrons is, however, as a rule
nearly always Maxwellian.  In this  low-density limit we know that the
level distributions are given by,
$$
\frac{N_i^k}{N_i^j} = n_e \frac{\langle v \sigma_{jk} \rangle}{A_k}
$$
where $N_i^k$  denotes the  number density of  the  species $i$  in its
level  $k$, $  v\sigma_{jk}  $  the  rate coefficient  for collisional
excitation from level $j$ up to level $k$, $A_k $ the total probability for
spontaneous transition from all higher levels down to $k$.
This approximation is applicable, if,
$$ 
n_e \ll \frac{A_k}{\langle v \sigma_{jk} \rangle}
$$

One important assumption here is that collisional excitations outweigh
radiative excitations which  is always true  as long as there are only
moderate external radiation fluxes.  For the s--p excitation in Helium
one finds that the  coronal  limit holds  up to electron  densities of
$\sim 10^{17}$cm$^{-3}$. (  Here   we  used $  \langle  v  \sigma_{jk}
\rangle \sim 6 \times 10^{-9} \cm^3\s^{-1}$ at 400K  from Janev et al.
1987, (2.3.1),  and $A_k =  1.8 \times 10^9  \s^{-1}$ (Sobelman et al.
1979, p297)).  Thus the coronal limit  is valid for will be applicable
in all our intended  applications concerning structure  formation.  In
proto-galactic clouds for example, we expect total number densities of
about $100\cm^{-3}$.  For us the most  important point is that we find
nearly  every atom in  its ground state.   Thus it is not necessary to
treat multilevel atoms in our case.

Comparing the  time  scales  for collisional  excitation,  collisional
de-excitation (Lepp   and Shull,    1983)  and  also  the   transition
probabilities (Allison and Dalgarno,  1970) of molecular hydrogen, one
finds that for low densities  ($n_H < 10^4$cm$^{-3}$ the population of
excited states of  H$_2$ is by many orders  of magnitude smaller, than
the ground state.  This is a fact which  can also be  seen in Lepp and
Shull 1983, as well  as in the  work of Dove et  al.  1987 where  they
studied  dissociation of molecular  para-hydrogen  by  collisions with
helium.   They found that  at   low  densities the dissociation   rate
coefficient approaches a constant value,  which corresponds to  direct
collisional dissociation  out of  the ($v=0,  \  J=0$) level   only. A
strong radiation field in the UV is capable of populating the exciting
states and  hence changing the  chemical behaviour of \HH.  However, a
detailed study (Shull 1978) shows  that  only for fluxes greater  than
$\sim  5  \tento{-15} \erg   \ \cm^{-2} \   \s^{-1}  \ {\rm Hz}^{-1} \
\sr^{-1}$   UV  Pumping becomes effective.   We  conclude  that in our
intended calculations  we can assume  all our considered species to be
in their ground state.

\subsection{Selection of Reactions and Rate Coefficients}\label{sec:SelRC}

        Although primordial  gas is a  simple mixture  of hydrogen and
helium,  the ongoing physical reactions   in it are immense.  E.g.  in
Janev \etal 1987 one  finds more than 70  reactions only involving the
ground states of our species. It is obvious that in order to construct
a computationally feasible model that some selections have to be made.
The quest is to make them as good as possible i.e.  without neglecting
any important  physics.  To   neglect  a collisional  processes   one,
obviously, has to make sure that
\begin{enumerate}
\item for all reactants and products the rate will for any temperature
and density never dominate the right hand side of the rate equations.
\item the reaction enthalpy is always negligible energy contribution
or loss to the internal energy of the gas.
\end{enumerate}
We used   these criteria to  construct our   reaction network out  of
hundreds of reactions found in the literature as well as in databases.

\section{Collisional Processes}

  In  appendix   \ref{app:rates}, we  present all  included processes,
their rate coefficients  and the according reference.  Discussing  the
reliability of the used rate   coefficients and their sources will  be
the main  aim of this  section.   Further information on atomic rates
can be found at
\href{http://www.pa.uky.edu/~verner/atom.html}{
Dima Verners 1996 Atomic Data Page}

\subsection{Ionizing processes}\label{sec:collisionalionization}

For the cosmological problems we are interested in, we have seen above
that  spontaneous decay is   a lot faster than  collisional excitation
(coronal  limit).  This already pointed out   the non-LTE character of
our  problem.  Metaphorically speaking, the  radiation field is in our
case much smaller than  in the LTE case,  since  there are not  enough
collisional excitations to build up  a Planckian spectrum.  Therefore,
if we compare in the following collisional and radiative processes, we
can use arguments derived for strict LTE for qualitative statements on
non-LTE conditions by assuming that the  radiation temperature is much
less than the  matter temperature.  This is  always true in the matter
dominated phase of the universe, when radiation  fields other than the
CBR are negligible.  We  adopt here Mihalas'  (1978, p123) estimate of
the ratio  between the total  number  of photoionizations $R_{ik}$ and
the total number of collisional ionizations $C_{ik}$,
$$
\frac{R_{ik}}{C_{ik}} \approx  \frac{4 (2 \pi^3 )^{\frac{1}{2}} E_i^3}
                        {3 m^{\frac{1}{2}} e^2 h^2 c^3} \left( \frac{W
                        k_B  T_R}{n_e (k_B T_e)^{\frac{1}{2}}} \right)
                        \exp \left[ h \nu_0 \left( \frac{1}{k_B T_e} -
                        \frac{1}{k_B T_R} \right) \right],
$$
where $E_i$ is the ionization energy out of the level  $i$, $k_B$ is the
Boltzmann constant, $m$  is the   electron  mass, $T_e$ the   electron
temperature, $e$ the electron charge, $c$  the speed of light, and $W$
is  defined   by $J_{\nu}  = W   B_{\nu}(T_R)$, where  $J_{\nu}$ is the
specific intensity at the  frequency $\nu$, and $B_{\nu}(T_R)$ denotes
the Planck spectrum of a black body with temperature, $T_R$.

This  gives  for ionization  of   hydrogen out  of the   ground  state
($E_i=13.6  $eV)  with   the   number  density of  free   electrons in
cm$^{-3}$,
$$
\frac{R_{ik}}{C_{ik}} \approx 1.55 \times 10^{14} \frac{W}{n_e} \left(
      \frac{T_R^2}{T_e}  \right)      ^{\frac{1}{2}}  e^{157820 \left(
      \frac{1}{T_e} - \frac{1}{T_R} \right)}
$$ 
In thermal  equilibrium with  $T_e =  T_R \sim 10^4$   and $n_e \sim 1
$cm$^{-3}$ this ratio is of  the order $10^{15}$, therefore  radiative
ionization  dominates by far. For $T_e  \sim 10 \times T_R \sim 10^3$,
however, the ratio  $\frac{R_{ik}}{C_{ik}}$ is of the order $10^{-47}$
and  we  see that  for all  of   our intended applications collisional
ionization is the dominating  process.   It is important, however,  to
include photoionization to  study the influence  of external radiation
fields on   the chemistry and the cooling   to heating balance  of the
structure forming gas, as, e.g., Lyman Alpha clouds in the vicinity of
quasars which are observationally accessible.

\paragraph{(1) Collisional Ionization of Hydrogen}\label{reac:1}

We use the fit  of Janev et  al. (1987, 2.1.5).  For temperatures below
$1000$K$ \sim 0.086 $eV we  take it to be zero  since it is smaller than
$10^{-60}$cm$^3$s$^{-1}$. Since in our  simulations we do not expect  to
encounter temperatures above $10^5$eV no relativistic corrections were
made. The ionization threshold is $13.6$eV. 

\paragraph{(2) Collisional Ionization of Helium}\label{reac:2}

We use the  fit given by Janev et  al. (1987,  2.3.9) For temperatures
below  $4000$K we  take it  to  be zero   since   it is smaller   than
$10^{-38}$cm$^3$s$^{-1}$. The threshold lies at $24.6$eV.

\paragraph{(3) Collisional Ionization of He$^+$}\label{reac:3}

The   rate  is  given  in      the  Aladdin Database (1989)   of    the
IEADS (International Atomic Energy  Agency, Data Section).   The energy
threshold is $54.4 $eV.  Again we only have to consider ionization out
of the ground state and no relativistic effects.

\subsection{Radiative Recombination}
\

Radiative  recombination  is the inverse reaction  of photoionization.
With a  similar argument  as the  one we  used to discuss the
relative importance of  collisional to radiative ionization, one finds
(Mihalas  1978)   that radiative  recombination  always  outweighs the
collisional  one.  This is especially  valid in  our low-density limit
since  the collisional   recombination   which is, as   all  three-body
processes, typically negligible for densities  smaller than about $10^8
\cm^{-3}$.

\paragraph{(4) Radiative Recombination to Hydrogen}\label{reac:4}

In the coronal limit we assume that recombination  can happen into any
quantum  state which, if  it  is an  excited state, will spontaneously
decay to the  ground state. Therefore, the  total  rate coefficient is
the sum of the   rate coefficients $\alpha_n$ for  all  $n=1..\infty$.
Ferland et al. (1992)  computed hydrogenic rate coefficients which are
in principle exact.  We  have fitted their  data (the sum of  all rate
coefficients for $n=1..1000$)  to a form  similar to what Janev et al.
(1987)  used  in  their compendium.   We made  sure  that  the fit is
accurate for temperatures from $1$K to $10^9$K.

\paragraph{(5, 6) Recombination to Helium}\label{reac:5}\label{reac:6}

He$^+$  is the   only  species in our  model,   which  is  subject  to
di-electronic recombination which dominates at high temperatures ($T >
6 \times 10^4$K).  Since   radiative and di-electronic   recombination
rates  are  independent  of  density  their   sum  gives  the  overall
recombination rate. For the  radiative recombination process we employ
the  rate coefficient  given by  Cen (1992)  and for the di-electronic
recombination the one given by Aldrovandi \& Pequignot (1973).

\paragraph{(7) Photo-attachment of Hydrogen}\label{reac:7}

The rate coefficient from Hutchins (1976) is  stated to be accurate to
10\% in the temperature range $100{\rm K}(=0.0086 {\rm eV})\leq T \leq
2500{\rm K}(=0.254{\rm   eV})$.  
The  cross  section for  the inverse  reaction  has been calculated by
Wishart (1979)    to   within 1\%    around  the  threshold  ($2\times
10^{14}$Hz$<\nu <2  \times 10^{15}$Hz).  From  that data alone one can
compute the rate for photo-attachment  from $2000$K to $10000$K, using
the principle  of detailed  balance or  simply the  Saha equation.  To
cover  a greater temperature range, however,  we use the cross section given
in de Jong (1972) for all frequencies outside of the interval given by
Wishart (1979). With that we are able to compute  and fit this rate in
the temperature range from $1$K to $10^8$K, with an accuracy to within
a  few  percent   for  $1$K   $<T<100$K,    about one  percent     for
$100$K$<T<10^4$K, and better than 10\% for $T>10^4$K.  At LTE the Saha
equation has to be valid and the rate at non-LTE will be naturally the
same as  in LTE since it  is an atomic property.  The Saha-Boltzmann
ionization equation reads,
\begin{eqnarray*}
\frac{k_{det}}{k_{att}}  =    \frac{N_{r+1}}{N_r}        N_e         =
\frac{u_{r+1}}{u_r}  \frac{2  (2  \pi m    k  T)^{3/2} }{h^3}   \exp(-
\frac{I_r}{kT}),
\end{eqnarray*}
where $N_e$ denotes the free  electron  number density, $N_{r+1}$  the
number density of neutral hydrogen  atoms, $N_r$ the number denisty of
H$^-$ ions,  $k_{det}$   the  rate coefficient  for  photo-detachment,
$k_{att}$ the rate  coefficient for photo-attachment, $u_r$, $u_{r+1}$
the respective partition     functions,  $I_r$  the  threshold     for
photo-detachment of H$^-$ ($0.755$eV), $m$ the  electron mass, and $k$
the Boltzmann  constant. The single bound state  of H$^-$ is $^1$S and
the ground state of H  is $^2$S.  Hence $u_{r+1}  = 2$ and $u_r =  1$.
The  rate  coefficient,  $k_{det}$,  for   photo-attachment in  LTE is
derived  by integrating the cross  section over a Planckian black body
spectrum. The  fit, which  is given in   the appendix, equals  the one
given by  Hutchins (1976) to within a  few percent in the  range where
the latter is applicable.

\paragraph{(8) Formation of molecular Hydrogen via H$^-$}\label{reac:8}

The rate coefficient has been taken from  Janev et al. (1987, 7.3.2.b)
and is based  on  theoretical cross  sections of  Browne and  Dalgarno
(1969)  and normalized   to  the  experimental  results  of  Hummer et
al. (1960) in the region of $500$eV to $4 \times 10^4$eV.  We averaged
their energy dependent rate   coefficient over a  Maxwellian  velocity
distribution of the incident particle.  The used data is only reliable
for  temperatures  above $0.1$eV.   However,  we,  know that  the rate
coefficient is  constant for small temperatures  and we also  see that
the energy dependent rate coefficient is already  the same for $0.1$eV
and    $1$eV   of the  incident    particle   for   a  temperature  of
$0.1$eV. Therefore, we take the rate coefficient at $0.01$eV to be the
same  as at $0.1$eV,   fit it, and set  it   to be constant  for lower
values.    This leads  to a  value  of the   rate coefficient for  low
temperatures  about  10\% higher  than the  constant   values given by
Bieniek (1980) and  de Jong (1972).   Launay, Le Dourneuf, and Zeippen
(1991) computed the ratecoefficient for this reaction depdening on the
ro-vibrational  state of    the  produced \HH    molecule.   Using the
potential derived  by Senekowitsch \etal they found  a thermal rate at
$300\K$ of  $1.49 \times  10^{-9}$ which lies  4\% above  the rate  we
derived from the data of Janev et al. (1987).

\paragraph{(9) Formation of H$_2^+$}\label{reac:9}

The rate coefficient for this radiative  association reaction has been
calculated by  Ramaker and  Peek  (1976) and has  then  been fitted by
Shapiro and Kang  (1987)  as well as  by  Rawlings et al.  (1993).  We
prefer the fit given by Shapiro and  Kang, because it covers a greater
temperature interval.

\paragraph{(10) H$_2$ Formation via  H$_2^+$}\label{reac:10}

\begin{figure}[hb]
\centerline{\psfig{file=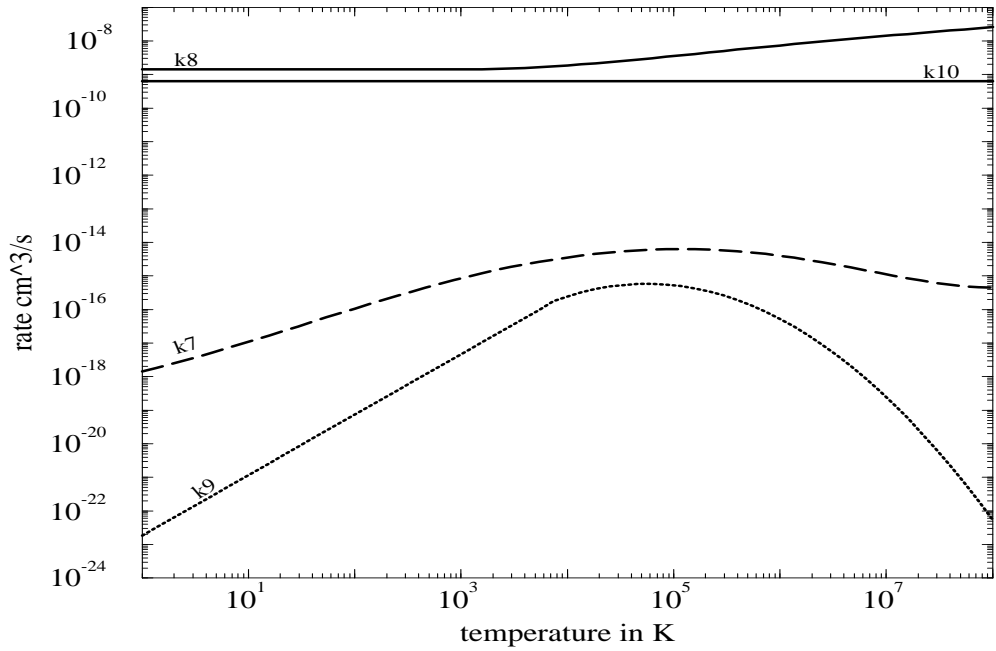,height=10cm,width=15cm}}
\caption{The  rate  coefficients    for  photo-attachment to
  H$^-$ (dashed line), H$_2^+$  formation  (dotted line) and the   two
  H$_2$ forming reactions, proc.  (8) and (10) (solid lines).}
\label{Fig:H2Form}
\end{figure}

The rate constant for  this charge  exchange reaction between  neutral
hydrogen and  H$_2^+$  was measured by  Karpas  et  al.  (1979) to  an
accuracy   better than 20\%.  This  rate  is constant  at low energies
because the distribution of final states in  phase space is determined
by the energy released in the reaction, which is almost independent of
the incident energy (Peebles 1993).  By assuming it  to be constant at
high temperatures as  well, we potentially introduce  an  error in the
H$_2$ abundance, because  the  by far dominating  destroying processes
will be balanced by a wrong formation  rate.  However, this error will
not  be  significant because the  formation of   hydrogen molecules is
dominated by the H$^-$ formation path.
%
% figure \ref{Fig:H2Form} goes here
%
Figure \ref{Fig:H2Form} summarizes the rate coefficients for formation
of H$^-$ and H$_2^+$ as well as the ones for H$_2$.  

\paragraph{(11) Charge Exchange between Molecular Hydrogen and
H$^+$}\label{reac:11}

This is the inverse reaction of the above. The rate coefficient was
derived by Donahue and Shull using detailed balance and the Karpas et
al. (1979) data to
\begin{eqnarray*}
k_{11} = 6.4 \times 10^{-10} \exp\left( -\frac{2.65 {\rm eV}}{T_{eV}} \right).
\end{eqnarray*}
That rate differs drastically  from the one  we have computed from data
given  in   Janev et al. (1987)    which   is accurate  also for   high
temperatures. The data  of Janev \etal  has  to be considered by  far
more accurate than the simple detailed balance argument of Donahue and
Shull (1991). However,  comparing results of our model  with the one of
Shapiro and  Kang (1987) we find  that drastic differences in this rate
do not change the final abundances significantly. 

\paragraph{(12) Dissociation of molecular Hydrogen by Electrons}\label{reac:12}

We  employ the rate coefficient that   Donahue and Shull (1991) derived
from the cross section given by Corrigan (1965).

\paragraph{(13) Dissociation of Molecular Hydrogen by neutral
Hydrogen} \label{reac:13}

Dove and Mandy  used fully 3D quasi-classical trajectory calculations,
using  the Siegbahn-Liu-Truhlar-Horowitz  potential energy surface, to
calculate the rate coefficient in  the low density limit.  They  found
significant  differences to the   former ``standard'' values  given by
Lepp  and Shull  (1983).   We use  the  results   for  the total  rate
coefficient of Dove  and  Mandy  (1986),  which assumes that  all  the
quasi-bound states will dissociate.  Although their ratecoefficient is
not accurate at temperatures well below $500\K$ this does not lead to a
significant error in the \HH abundance as can clearly be seen from the
discussion in section \ref{sec:H2chem}

\paragraph{(14) Collisional Detachment of H$^-$ by Electrons}\label{reac:14}

We take  the rate coefficient from Janev  et al. 1987 (7.1.1),  which is
based on experimental results. The net energy loss is as in ionization
reactions the threshold energy which is $0.755$eV.

\paragraph{(15) Collisional Electron Detachment of H$^-$ by neutral
Hydrogen}\label{reac:15}

The formation  of  an auto-dissociating state of   molecular hydrogen,
(equation    \ref{eq:autodiss}  in section  \ref{NeglectedHminusDiss})
seems to be  far more effective  in destroying  H$^-$ than  the direct
collisional detachment
\begin{displaymath}
\rm
H^- \ \ + \ \ H \ \ \rightarrow \ \ H \ \  + \ \ H \ \ +  \ \ e^-.
\end{displaymath}

The rate coefficient for  the auto-dissociating state below $0.1$eV is
very uncertain.  We have  extrapolated the data given  by Janev et al.
(1987) linearly for those  temperatures.  This overestimates  the rate
coefficient in  this   range.   However, it  does    not introduce any
significant error   in the abundances since  for  a collision of H$^-$
with  H in this temperature regime  it is far more  probable to form a
stable H$_2$ molecule  than to destroy  the H$^-$. In fact, we  expect
this reaction to be only  of importance in  a very small range  around
$1$eV, since  for   higher  temperatures,  the dissociation  by   free
electrons is far more effective than the one by H.

\paragraph{(16) Mutual Neutralization between H$^-$ and H$^+$}\label{reac:16}

The rate coefficient given  by Dalgarno  and  Lepp (1987)  is consistent
with the measurements by Szucs  \etal (1984)   and Peart, Bennett  and
Dolder (1985).    See the   latter   reference for   a  review  on   the
experimental difficulties and uncertainties of ion-ion collisions.

\paragraph{(17) H$_2^+$ Formation in H$^-$ and H$^+$
Collision}\label{reac:17}

Shapiro and  Kang (1987) derived  the rate coefficient from  the cross
section which was measured by Poulaert et al. (1978) in the range from
$0.001$eV to $3$eV and found to have a $E^{-0.9}$ dependence with $1.5
\times  10^{-14}$cm$^2$  at $0.003$eV.  The fit  given  in Shapiro and
Kangs   (1987), however, is discontinuous  at   $10^4$K. We changed it
within  4\%, which  surely   lies  within  the uncertainties   of  the
experimental data  to make it continuous.  The result is  given in the
appendix.

\paragraph{(18) Dissociative Recombination of H$_2^+$ with slow
Electrons}\label{reac:18}

The rate  for this reaction  was calculated using multichannel quantum
defect theory (MQDT) by Nakashima  et al. (1987)   in the energy  range
from $0.02$ to about $1$eV.
Their result  is much  smaller  at low  temperatures,  than previously
assumed (see  e.g. Shapiro and  Kang  1987, Mitchell and  Deveau 1983).
More   recently  Schneider    et   al. (1994)  recalculated  this  rate
coefficient  also  using  MQDT. Their tabulated  rate  coefficient for
dissociative recombination   of  the H$_2^+$   ground  state via   the
$\Sigma_g^+$ state is, above 1000K, by a factor of two to four smaller
than the rough fit  given by Nakashima.  Our  fit of the Schneider  et
al. (1994) data  neglects  the fact  that the   rate is  increasing for
$T<50$K.  This  is reasonable since one does  not expect this reaction
to be  important  for temperatures smaller than   $100$K due the small
abundance of free electrons, as well as H$_2^+$.

\paragraph{(19) Neutralization between H$_2^+$ and
H$^-$}\label{reac:19}

This  mutual  neutralization  reaction  takes place  at  low  energies
($<1$eV)   where H$_2^+$ and   H$^-$ coexist.  Unfortunately, the only
three   measurements we are aware   of, Aberth \etal   in Moseley et
al. (1975), Szucs \etal (1983), and Dolder \& Peart (1985), give only
data for energies  above $3$eV. We use the  rate  coefficient given by
Dalgarno  and  Lepp (1987)  despite  its uncertainty.  This equals the
coefficient  given by  Prasad and Huntress  (1980) which  was  used by
Shapiro and Kang (1987) to within 25\%.

\section{Photoionization and Dissociation Processes}\label{sec:PhotoProc}

In general we will only consider photoionization (dissociation) out of
the ground state since we assume the abundance of excited states to be
negligible.  The  rate of  a particular photoionization (dissociation)
reaction is given by:

\begin{eqnarray}\label{equ:photorate}
\frac{\partial \rho_k}{\partial t} / {\rho_k} =  \int^{\infty}_{\nu_{th}}4\pi 
    \sigma_\nu^k \frac{i(\nu)}{h\nu} d\nu,
\end{eqnarray}
where  $k$  denotes H, He$^+$,   H$^-$, H$_2^+$ and  H$_2$
respectively,     $i(\nu)$ is the   intensity  of the radiation field,
$\nu_{th}$      the    threshold      energy     for    which    photo
ionization(dissociation)  is possible,and $\sigma_\nu^k$ the frequency
dependent photoionization (dissociation) cross section of species k.

\paragraph{(20 -- 22) Photoionization of H, He and
He$^+$}\label{reac:20}\label{reac:22}

Hydrogenic cross  sections have been  studied  in great detail.  Since
they are  relatively simple to   calculate the available data is  very
accurate.  We use  the typical expression  given in Osterbrook (1989).
The threshold  frequencies are $h\nu = 13.6  $eV, $24.6 $eV, and $54.4
$eV, for H, He, and \Hep, respectively.

\paragraph{(23) Photo-detachment of the H$^-$ Ion}\label{reac:23}

The best data   available is given by   Wishart (1979) with a   stated
accuracy within 1\%.  The fitting formula  for the cross section given
in Shapiro and  Kang (1987) is  accurate to $\sim   10\%$ at the  high
energy end. This is a  normal characteristic of such fitting formulas.
Although is  adequate for  present  purposes, we recommend   using the
tabulated values  of Wishart  (1979),  where they  are avaialabel,  in
order to derive the rate of H$^-$ destruction  since using the fitting
formula   in the integral of  equ.    \ref{equ:photorate} leads to  an
overestimation of the  according rate.  \Hm gets destroyed efficiently
by the CBR to redshifts  $\sim 110$.  A  fit for $k_{23}$ depending on
radiation     temperature        can         be         found       in
\href{http://www.mpa-garching.mpg.de/~max/minmass.html}{Tegmark  \etal
1996}.

\paragraph{(24) Photoionization of Molecular Hydrogen}\label{reac:24}

The high threshold of $15.42 $eV assures that the CBR  will not be able
to photo-ionize H$_2$ in the post-recombination universe. The rate is
taken from Shapiro and Kang (1987).

\paragraph{(25 -- 26)Photodissociation of
H$_2^+$}\label{reac:25}\label{reac:26} 
\ 

There   is a conceptual  difficulty in  the treatment  of this process
since for strong radiation  fields and high densities,  excited states
of H$_2^+$ will  be  populated.  For them the  photodissociation cross
sections  are  much higher  than  for the ground  state.  However, the
levels  might not be  in  their thermal  equilibirum distribution.  To
treat this exact, one  would    have to  include all H$_2^+$    levels
explicitly, which  for our applications would not  be worth the effort
since  most of the H$_2$   is formed via  H$^-$.   By using solely the
cross  section for the ground  state we will  overestimate the H$_2^+$
abundance.  Using the   LTE rate one could   derive a lower limit  and
estimate the error made, which we did not  attempt in our simulations,
yet.  The most recent data for dissociation out of the ground state as
well as LTE rates can be found in  Stancil (1994).  We have fitted the
cross section for  the ground state to  better than 2\%.  This fit  is
more   accurate,  and also does  not    show the unfortunate divergent
character  than the one  given  in Shapiro  and  Kang (1987) which was
drawn from Dunn (1968). The threshold energy is $2.65$eV.  For photons
with energies greater than   $30$eV, H$_2^+$ can get  dissociated with
two protons  as products  (proc.  26). For   this we adopted  the rate
given in Shapiro and Kang (1987).

\paragraph{(27) Photodissociation of H$_2$ by
predissociation}\label{reac:27}

Photodissociation   of   the  ground   state  of   molecular  hydrogen
($X^1\Sigma_g^+(v))$    happens   mostly through  absorption   in  the
Lyman-Werner Bands  to  the electronically  and  vibrationaly  excited
states, $B^1\Sigma_u^+(v')$ and $C^1\Pi_u(v')$ which then decay to the
continuum  of the ground state.   This is called  the two-step Solomon
process (cf. Stecher and Williams, 1967).  Allison and Dalgarno (1970)
computed the  band  oscillator  strengths and   Dalgarno  and Stephens
(1970)  derived the probability   that  those  states decay  into  the
continuum of the  ground state.  The dominant  photodissociation paths
are through  absorption in the   Lyman Band to  the vibrational states
$6<v'<20$. This means that H$_2$ dissociation happens mostly in a very
narrow frequency range  $12.24eV  <  h\nu  < 13.51$eV.  Assuming   the
incident   radiation  field  to    be   practically  constant   around
$h\overline{\nu}=12.87$eV,    which     corresponds  to  $v'=13$,  and
neglecting  self-shielding   we   can  derive   a rate   constant  for
photodissociation of H$_2$ through
\begin{eqnarray*}
k_{27} &=&   \sum_{v'}\frac{\pi  e^2}{mc}  f_0^{v'}   p_{Ly}^{v'} \int
         \frac{j(\nu)}{h\nu}       \phi_{v'}(\nu)       d\nu   \approx
         \frac{j(\overline{\nu})}{h\overline{\nu}}\sum_{v'}  \frac{\pi
         e^2}{mc}     f_0^{v'}  p_{Ly}^{v'}    \\
 &\sim&   1.1  \times 10^8
         \frac{j(\overline{\nu})}{{\rm ergs}{\rm Hz}^{-1}{\rm
s}^{-1}{\rm cm}^{-2}}\ s^{-1} ,
\end{eqnarray*}
where   $j$ denotes    the      radiation flux   in     s$^{-1}$ergs$\
$cm$^{-2}$Hz$^{-1}$  at $h\overline{\nu}=12.87$eV, $f_0^{v'}$      the
oscillator  strength   for  the   transition   $X^1\Sigma_g^+(v=0)$ to
$B^1\Sigma_u^+(v')$,   $   p_{Ly}^{v'}$    the      probability   that
$B^1\Sigma_u^+(v')$ decays to the continuum, $\phi_{v'}(\nu)$ the line
profile, and  $\pi e^2/(mc)= 2.65  \times 10^{-2}$cm$^2$ the classical
total cross section.

\paragraph{(28) Direct Photodissociation of \HH}\label{reac:28}

Direct  photodissociation of the ground state  by absorbtion into the
continua of the Lyman  and Werner systems  has been studied  by Allison
and Dalgarno 1969. We have fitted their cross sections within the
stated accuracy of their data with,
\begin{eqnarray}
\sigma_{28} = \frac{1}{y+1} (\sigma_{28}^{L0} + \sigma_{28}^{W0}) + 
   (1-\frac{1}{y+1}) (\sigma_{28}^{L1} + \sigma_{28}^{W1})
\end{eqnarray}
\noindent
where
\begin{eqnarray}
\sigma_{28}^{L0} & = & 10^{-18} \times  \left\{ \begin{array}{ll}
       \dex(15.1289 - 1.05139 \times h\nu) \ \cm^2, & {\rm \ \ \ }
         14.675\eV < h\nu < 16.820\eV   \\
       \dex(-31.41 + 1.8042 \tento{-2} (h\nu)^3 - &
4.2339\tento{-5} (h\nu)^5) \ \cm^2,  \\  {\rm \ \ \ }
        &  \ \ \ 16.820 \eV < h\nu < 17.6 \eV, 
              \end{array}
      \right. \\
\sigma_{28}^{W0} & = & 10^{-18} \times  \left\{ \begin{array}{ll}
       \dex(13.5311 - 0.9182618 h\nu) \ \cm^2, & {\rm \ \ \ }
         14.675\eV < h\nu < 17.7 \eV, \\
              \end{array}
      \right. \\
\sigma_{28}^{L1} & = & 10^{-18} \times  \left\{ \begin{array}{ll}
       \dex(12.0218406 - 0.819429 h\nu) \ \cm^2, & {\rm \ \ \ }
         14.159\eV < h\nu < 15.302 \eV, \\
       \dex(16.04644   - 1.082438 h\nu) \ \cm^2, &  {\rm \ \ \ }
         15.302 \eV < h\nu < 17.2 \eV, \\
              \end{array}
      \right. \\
\sigma_{28}^{W1} & = & 10^{-18} \times  \left\{ \begin{array}{ll}
       \dex(12.87367 - 0.85088597 h\nu) \ \cm^2, & {\rm \ \ \ }
         14.159\eV < h\nu < 17.2 \eV, 
              \end{array}
      \right.,
\end{eqnarray}
and   $y$ is the ortho-,  to  para-\HH ratio. Since unfortunately they
only gave the data  for photon excess energies up  to $3\eV$ one might
make quite  significant errors in   deriving the ratecoefficient where
one integrates over this cross section. The error  made depends on the
shape of the radiation spectrum.

\section{Brief Summary of Neglected Processes}\label{sec:Neglected}

In this section we will  briefly summarize some  processes which we find
to be negligible.   Although we  can  almost be sure that  this smmary
will be far from complete, we hope to  stimulate with this a discussion
which  should   lead  us  to a   more   detailed understanding of  the
primordial  gas  chemistry and enable  us to  find the minimum  set of
reactions,  which describes primordial  gas  accurately enough for the
desired cosmological applications.

\subsection{Photoionization Heating and Secondary Ionization}

For all  simulations where  we do want   to  consider strong  external
radiation fluxes we  have to look at  the fate of the  photo electrons
produced in  photoionizations. To  estimate the  importance of  this
effect we  first derive  the numbers of   high energy photo  electrons
produced in a typical quasar radiation field.

Shapiro and Kang (1987) modeled a quasar radiation flux with:
\begin{displaymath}\label{equ:J21}
F_\nu  =  1.0  \times  10^{-21}  \epsilon  \left(  \frac{\nu}{\nu_{H}}
   \right)^{-\alpha} ergs \ cm^{-2} \ s^{-1} \ Hz^{-1},
\end{displaymath} 
where $\nu_H$  is the  Lyman edge frequency,  $\alpha =  0.7$, $0.7eV<
h\nu \leq 12.4$keV, and $\epsilon$ is a parameter introduced to adjust
the amplitude of the flux (or effectively the distance of the quasar).
Although Sargent, Steidel, and Boksenberg  (1989) argued that  $\alpha
\geq 1$  might be  a better fit,   we  will derive an  upper  limit of
produced photo  electrons  with the  more drastic value  of $\alpha  =
0.7$.   We calculated the fraction $n_{\gamma,\ e^-}^{low}/n_{\gamma,\
e^-}^{high}$  between the electrons produced   by photons in the range
from  $h\nu_H$ to $2h\nu_H$ and  the  ones produced by  photons in the
range of $2h\nu_H$ to $12.4$keV:
\begin{displaymath}
\frac{n_{\gamma,\ e^-}^{low}}{n_{\gamma,\ e^-}^{high}} =
 \frac{\int_{\nu_H}^{2\nu_H} F_{\nu} \frac{\sigma_{20}(\nu)}{h\nu} d\nu}
      {\int_{2\nu_H}^{24keV/h} F_{\nu} \frac{\sigma_{20}(\nu)}{h\nu} d\nu}.
\end{displaymath}
For a  very  hard spectrum with   $\alpha =  0.7$  we get $n_{\gamma,\
e^-}^{low}/n_{\gamma,\ e^-}^{high}  = 10$ whereas    for a more   soft
spectrum this number  increases (e.g.  13  for $\alpha=1.0$).  Because
photo electrons have an energy of  $h(\nu - \nu_H)$, the ones produced
by photons in the range from $h\nu_H$ to $2h\nu_H$ will not be capable
of collisionally ionizing  any species other   than H$^-$.  Therefore,
all they can do is  to either go  into heat, which means scatter  with
other electrons or ions and  equilibrate to a Maxwellian distribution,
or  excite other atoms, ions or  molecules.  The photo electrons above
$2h\nu_H$  carry  enough  energy  to also   ionize  (dissociate) other
species.  The very  high   energetic  ones are capable   of   ionizing
(dissociating) many atoms  (molecules) while cooling down.  Shull  and
van Steenberg  (1985) gave fits  for the  fraction of  photo electrons
above    $100$eV which go     into  heat, ionization,  and  excitation
respectively.  We  however  find that the  total energy  of all  photo
electrons above $100$eV is  only 1.5\% of the total  energy of all the
photo  electrons produced in  photoionization of neutral hydrogen by a
typical quasar flux.  Hence we do not believe those photo electrons to
be significant.  The ones in the range  of $h\nu_{th}$ to $100$eV make
up about one  third of all of the  produced photo electrons.  Fig.3 of
Shull  and Van Steenberg  (1985) shows that  even for a mostly neutral
gas, maximally about 30\% secondary electrons will go into ionization.
We conclude that  by leaving out secondary  ionization we will make an
error  in the estimates of  the ionized  fraction  of always less than
10\%.

The estimates above were based on the photoionization cross section of
hydrogen, but they hold in general for the  following reasons.  First,
the cross sections for He  and He$^+$ scale  as $\nu^{-3}$ in the high
frequency limit,  and  second, nucleosynthesis predicts $n_{H}\sim  16
n_{He}$. Therefore the photo electrons produced in photoionizations of
neutral hydrogen are the dominant ones.  Clearly, the above discussion
only justifies  the approximation of  leaving out the effects of photo
electrons,   for   applications    where  the    radiation spectrum is
approximated well by a power law as given in equ. \ref{equ:J21}.

\subsection{Exciting Collisions}
 
For  our  intended  low density  applications   the coronal limit (see
above)  holds and we therefore know  that any  excitation of H, H$^-$,
He, He$^{+}$,  and H$_2$ will be followed  by a spontaneous decay back
into the ground state and so ensure that practically every atom or ion
will be in its ground state most of  the time.  An excitation does not
change the abundance of   our model species   and does not   enter the
reaction network explicitly.   We, however, will consider the dominant
ones in the energy balance through appropriate cooling functions.

\subsection{Collisions of Electrons with Atoms and Ions}

Electrons can photo-attach  with, ionize, and excite, neutral hydrogen
and helium and also ionize, recombine with, and excite their ions.  We
included    all  radiative  recombination  and collisional  ionization
reactions  (procs.~1--6, 14) except  the  double ionization of  helium
since it  does not significantly  influence the ionization balance due
to   its high threshold.   The  dominant  excitations  (see above) are
treated through the appropriate cooling  functions.  We also  included
the  photo--attachment  reaction of neutral   hydrogen (proc.~7).   No
reference  for photo--attachment of neutral  helium could be found but
one  can be confident that  its effects are  by far dominated by H$^-$
which is the main species for the  formation of molecular hydrogen and
thereby controlling the cooling of primordial gas at low temperatures.

\subsection{Proton Collisions with Neutral Helium and Hydrogen}

Below   temperatures of  $\sim 1$eV the   neutral  atoms are  the most
abundant species,   whereas free protons and He$^+$   are, even in the
non--equilibrium case, more than two orders of magnitude less abundant
than H,   and  He respectively  (Shapiro  and  Kang   1987).  For  the
equilibrium case this difference is far more drastic.

We checked all proton collisions listed in Janev et  al. 1987 and find
that the rate coefficients typically drop drastically for temperatures
below $10$eV. This  is especially true  for  collisional ionization by
protons. The exception to the rule are charge exchange reactions as,
\begin{displaymath}
\rm H(n) \ \ + p \ \ \rightarrow \rm \ \ p \ \ + \ \ H(n), \ \ \ \ n=1, 2, 3, ....
\end{displaymath}
Since we  expect nearly all hydrogen atoms  to be in the ground state,
the $n=1$ case will be  the most probable  process.  Although it  does
not  enter  the  reaction   network,  since   it  ``produces what   it
destroys'', it is an important reaction theoretically since it ensures
tight thermal and  spatial coupling between  the protons and the  ions
due  to   its relatively  high rate  coefficient  of   $\sim 10^{-8} \
\cm^3\s^{-1}$.

All  rate coefficients  for  the  excitation  of neutral hydrogen  and
helium by protons  out of the ground state  are smaller by many orders
of magnitude than the one  for excitation by electrons.  Since  charge
neutrality requires that there  are  about as  many free  electrons as
protons, we know that excitation by electrons dominates excitations by
protons.

\subsection{Collisions of Hydrogen with Helium and their Ions}

An interesting  feature of the  presented reaction network is that all
species which are build up by  hydrogen nuclei couple with the species
formed  by helium nuclei only by  the  free electron fraction. This is
due to the following reasons;
\begin{enumerate}
\item Electrons are much  more  effective in  ionizing  than any other
      species      since      their        mean      velocity       is
      $(Am_H/m_e)^{1/2}\approx43A^{1/2}$ times  higher than for an ion
      with $A$ times the proton mass.
\item Charge  Exchange between H  and  He$^+$ has  a  rate coefficient
      $\sim  1.9 \times 10^{-15}   \cm^3\s^{-1}$ and is according to
      Couchman  (1985) negligible.
\item H$^-$,   H$_2^+$, and H$_2$ are  not  capable of ionizing helium.
      They are relatively fragile   and have very
      small abundances at temperatures above $\sim$1eV where they have
      kinetic energies far below ionization thresholds.
\end{enumerate}
A discussion why the reaction network does not include processes where
helium or  its ions destroy H$^-$, H$_2^+$,  or H$_2$ can  be found in
the subsequent sections.

\subsection{Neglected Dissociating Reactions of H$^-$}
 \label{NeglectedHminusDiss}

Electron  detachment of H$^-$  through neutral helium has according to
the ALADDIN database of the IAEADS, a cross section of the order $\sim
10^{-16} \cm^2$ and is therefore by many orders smaller than the cross
section  ($\sim 10^{-9}-10^{-7} \cm^2$; Janev et  al. 1987, 7.3.2a) of
reaction
\begin{equation}\label{eq:autodiss}
H^- \ \ + \ \ H \ \  \rightarrow \ \  H_2^{-*}(\Sigma_g)\ \ 
                     \rightarrow \ \  H \ \   +\ \  H\ \ +\ \ e^-.
\end{equation} 
  Exactly the same argument holds   for electron detachment of   H$^-$
through molecular    hydrogen   which has   a    cross section   $\sim
10^{-18}-10^{-16} cm^2$ (ALADDIN database).

We did not  include mutual neutralization of H$^-$  with  He$^+$.  The
available data from the IAEADS via the ALADDIN database, which is only
valid for energies  above 42eV, indicates   that the cross  section of
mutual neutralization  of H$^-$ with He$^+$ is  higher than for mutual
neutralization with protons.  That  is  what one might naively  expect
when  considering that the   Coulomb attraction, acts  as  the driving
force for   this reaction, competing  with the  kinetic motion  of the
reactants.  Since  He$^+$ is heavier than H  its mean velocity will be
slower  but  the  Coulomb  attraction is   the  same  therefore mutual
neutralization  with He$^+$ should occur with  a higher rate.  We only
leave out this reaction due to the lack of reliable data. However, the
expected  error is small since at  low temperatures most  of the H$^-$
will react with  neutral hydrogen and  form  molecular hydrogen (proc.
7) and at  higher  temperatures ($>$8,000K) collisional detachment  by
free electrons (proc.  14) will dominate the destruction of H$^-$.

Neutralization between H$^-$ and He$^{++}$ can be neglected due to the
high  temperature   threshhold   for  He$^{++}$  formation.   Whenever
He$^{++}$ is formed  the temperature is so  high that nearly all H$^-$
will be destroyed.

\subsection{Omitted reactions for H$_2$ formation}

Molecular hydrogen formation by excited atom radiative association
\begin{displaymath}
\rm
H(n=2) \ \ + \ \ H(n=1) \ \ \rightarrow {\rm \ \ H}_2 \ \ + \ \ h\nu
\end{displaymath}
has  a very small rate  coefficient, due to  the zero dipole moment of
molecular hydrogen (Latter and Black 1991). Also, for the cosmological
applications  we are interested in, the  $n=2$ population is extremely
small in  all cosmological   applications  making this  process   less
significant    than  the    dominant    H$_2$  formation    mechanisms
(procs. 8,10).

We also neglect the formation path  where H$_2^+$ first gets formed by
excited atom  radiative association and  then forms molecular hydrogen
through the charge  exchange with neutral hydrogen (proc.10) (Rawlings
\etal  1993).
\begin{eqnarray*}
\rm
H(n=2) \ \ + \ \ H(n=1) \ \ & \rightarrow &\rm \ \ H_2^+ \ \  +  \ \ e^-,  \\
\rm
H_2^+  \ \ + \ \ H(n=1) \ \ & \rightarrow &\rm \ \ H_2   \ \  +  \ \ H^+.
\end{eqnarray*}
This process can in certain circumstances,  e.g.  dense and hot gas in
circumstellar environment,  dominate     formation  by  excited   atom
radiative association In  our intended cosmological applications these
processes, however, will never be significant, since the population of
excited states is always negligible.

\subsection{Collisional Dissociation of \HH and \HHp}\label{sec:CDofH2}

\begin{figure}[hb]
\centerline{\psfig{file=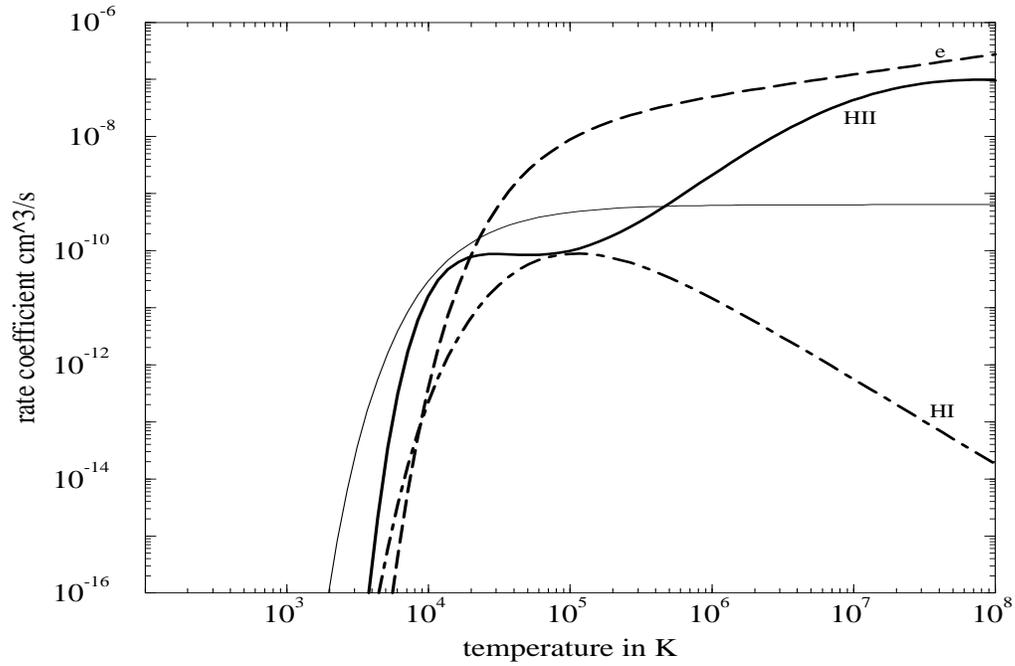,height=10cm,width=15cm}}
\caption{Rate  coefficients
  for  the dissociation   of  molecular hydrogen  by neutral  hydrogen
  (dot-dashed  line), and free electrons  (dashed  line). 
  The light solid   line   depicts the  rate coefficient   for  charge
  exchange between H$^+$ and H$_2$, given by Donahue and Shull (1991),
  whereas the  thick solid line is  our numerical integration from the
  Janev et al.  data.} \label{Fig:H2DestrRates}
\end{figure}

Since  the   hydrogen  molecule is    the main   coolant  for  gas  at
temperatures    below $\sim$1eV  we   pay  special  attention  to  its
destruction mechanisms.  We follow  here Shapiro  and Kang (1987)  and
assume    that  at every dissociation  $4.476$eV    are  lost.  In the
following we will  discuss dissociating reactions which  we find to be
negligible compared to    the three  dissociating reactions  we   have
included     in our model  and   which    are  illustrated in   Figure
\ref{Fig:H2DestrRates}.
%
% figure \ref{Fig:H2DestrRates} goes here
%

\subsubsection{Dissociation of molecular Hydrogen by $e^-$}

\paragraph*{\ \  i) H$_2^*$ + e$^-$ $\rightarrow$   (H$_2^-)^*$  
                                        $\rightarrow$ H$^-$  + H }
\

Wadehra and Bardsley  (1978) showed that  this dissociative attachment
reaction depends strongly on   the vibrational and  rotational states.
For low density gas ($n_H<10^5$cm$^{-3}$) in which nearly all hydrogen
molecules  will  be in their ground  state  its  rate is  of the order
$10^{-15}$cm$^3$s$^{-1}$ for an  electron  temperature of $1$eV.   For
electron temperatures below  that  this  rate drops drastically   even
further. After  comparing this   to Figure  \ref{Fig:H2DestrRates}  we
conclude that  this process will never  play an important role for the
destruction  of H$_2$ in our  applications.  It is, however, a crucial
process in situations with either densities  above $10^4 $cm$^{-3}$ or
intense   ultraviolet    radiation  fluxes,    because  under    those
circumstances    the excited    states   will  become  populated  (for
vibrational excitation of molecular  hydrogen in intense UV fluxes see
Shull 1978).

\paragraph*{\ \ ii) H$_2$ + $e^-$ $\rightarrow$ H$^*$ + H$^*$ + $e^-$}  
       \label{sec:H2Dissociationbye}, 

Our process 12 is,
\begin{eqnarray}\label{reac:H2byE1}
\rm
H_2 + e^- \rightarrow  H(1s) + H(1s) + e^-,
\end{eqnarray}
and dominates all other reactions of the type,
\begin{eqnarray}\label{reac:H2byE2}
\rm
H_2 + e^- \rightarrow H^* + H^* + e^-.
\end{eqnarray}
Intuitively    one   would    have    expected   this      because  in
proc. \ref{reac:H2byE2} the incident energy has to be enough to excite
the  produced H atoms. However,  the $2s$  level already lies $10.2$eV
above the ground   state which only  few  electrons  have at the   low
temperature ($< 1$eV) where molecular hydrogen exists.

From the graphs in Janev et al. 1987,  (2.2.5) it is evident that also
$$
H_2 + e^- \rightarrow H_2^- \left( \begin{array}{c}
                             b^3\Sigma_u^+      \\
                             a^3\Sigma_g^+      \\
                             c^3\Pi_u           
                             \end{array}
  \right) \rightarrow H(1s) + H(1s) + e^-
$$
will be   dominated by the  above reaction  \ref{reac:H2byE1}.  Please
note that the rate coefficient of reaction 2.2.6  in Janev et al. 1987
is a factor   10 to high!   (Personal communication with  Bill Langer,
1995)

\subsubsection{Dissociation of Molecular Hydrogen by H$^+$}

The  positive   hydrogen ion is able    to  destroy molecular hydrogen
through  the charge exchange  reaction  (11).  The direct  collisional
dissociation,
\begin{eqnarray*}
\rm
H^+ \ \ + \ \ H_2 \ \ \rightarrow \ \ 2H \ \ + H^+ ,
\end{eqnarray*} 
has often  been left out by  former studies. We also  were not able to
find a  rate  coefficient for  this  reaction. However,   one does not
believe this to be important since
\begin{enumerate}
\item at  low temperatures where  a  significant H$_2$ fraction  exists
      there  will  be   nearly  no  free protons    due  to the   high
      recombination rate at low temperatures.
\item dissociation by electrons is likely to be  more effective due to
      their higher velocities in the single temperature fluid.
\end{enumerate}

\subsubsection{Dissociation of Molecular Hydrogen by He}

Dove and Mandy  (1986b) found that  He  in comparison  to $H$ is  very
inefficient in  dissociating  H$_2(0,0)$.   They further state   that,
evidently  the collision between   two closed shell  species tends  to
cause the molecular bond to stiffen, whereas a  collision with an atom
weakens and loosens the molecular  bond. In a  follow up paper Dove et
al. (1987) applied their result  to  interstellar densities and  found
that the dissociation through   neutral helium is negligible and  that
H$_2$-He collisions do not excite  the vib./rot.  populations of H$_2$
in   low  density gas.    Following thess results   we do  not include
dissociation of H$_2(0,0)$ by He.

\subsubsection{Dissociation of Molecular Hydrogen by He$^+$}

The number fraction of He$^+$ in primordial gas  is about 10\%. Due to
the high  ionization threshold  of He$^+$  of $54.4$eV  it will not be
abundant at low temperatures.  Furthermore, from the scaling relations
of the  rate coefficients  given in  Janev  \etal (1987), we  know the
dissociation  of H$_2$ by He$^+$  occurs  $0.35$ times  less than  the
corresponding proton reaction.  Therefore  we can safely conclude that
this reaction can be omitted.

\subsubsection{Dissociation of Molecular Hydrogen by H$_2$}
     
    The dissociation  rate coefficient for  H$_2$ is of the same order
    and temperature dependence as the one  for dissociation by neutral
    hydrogen  (Lepp and Shull, 1983).  However  we do not include this
    reaction since the  \HH to H   fraction (and hence the rate)  is
    always smaller than $\sim 10^{-2}$ and therefore negligible.

The study  of Dove et  al. (1987a,b)  showed  that, in the low-density
limit, dissociation of molecular hydrogen by collisions with helium at
temperatures in the range $2000-10^4\K$ is  negligible compared to the
\HH  - H collisions.   Only for temperatures  close to $10^4\K$ do the
two rates  become of the  same order.  We  can, however, still neglect
this  process  since helium  has a much   smaller  number density than
hydrogen,  and  second  we  also expect  dissociation   via the charge
exchange reaction with \Hp to dominate at such high temperatures.

We  do  not  know the   rate for  direct  collisional  dissociation of
molecular hydrogen by protons. However, by simply  looking at the rate
coefficients for ionization of H$_2$  and dissociation of H$_2^+$  by
protons (Janev  et al.1987, 3.2.5,  3.2.6) one  sees that  those  rate
coefficients   drop drastically   for    temperatures  below   $10$eV.
Therefore,  we   are confident that    dissociation by protons is
negligible.   For  dissociation by  He$^{++}$   ions exactly  the same
arguments  used for  protons are valid  (see  Janev et al.  1987, 5.2.1,
5.2.3).  In addition the charge exchange reaction between H$_2$ and He$^{++}$
is highly inefficient for temperatures below $100$eV.

\subsection{Collisional induced Absorption}

Molecular hydrogen  does not have  an  electric  dipole moment in  its
ground electronic state.As a  result  absorption of photons  can  only
take place   via electric  quadrupole  transitions   and  low--density
molecular  hydrogen  gas  is  essentially  transparant  throughout the
visible and   infrared part of  the spectrum.   However,  each  time a
collision between  two particles occurs,  the interacting  pair (\HH -
\HH, \HH  - He, \ \HH -  H) acts as  a ``virtual supermolecule'' which
due to its    nonzero electric dipole,    can absorb photons   with  a
probability  which is   much higher   than  that  of an   isolated \HH
molecule.   However,  this process can,  according  to the  results of
Lenzuni, Chernoff, and  Salpeter (1991), only contribute significantly
to   the opacity for  number    densities greater than $\sim   10^{18}
\cm^{-3}$,  and hence is  negligible  at the low densities  considered
here.

\section{The Cooling and Heating Functions for Optical Thin Gas}

To model the cooling behaviour of the gas correctly, we incorporate the
following cooling and heating mechanisms,
\begin{itemize}
\item Compton cooling
\item Recombination cooling due to hydrogen and helium
\item Line cooling of hydrogen and helium 
\item Bremsstrahlung cooling
\item Molecular formation and line cooling
\item Photoionization heating
\item Photodissociation heating
\end{itemize}
The   cooling    functions  for  Compton,  recombination,   line,  and
bremsstrahlung cooling we have used are  given in AZAN96.  The molecular
cooling and heating rates are somewhat controversial in the literature
and, therefore, deserve special attention.

\subsection{Molecular Cooling}

\subparagraph{Formation cooling of molecular Hydrogen}
\ 

Most of the reaction enthalpy ($1.83\eV$ for process 10 and $ 3.53\eV$
for  process   8) goes rather  into   ro-vibrational excitation of the
molecule than to the  kinetic energy of  the  outgoing particles.   We
assume that  all  the reaction energy  goes  into excitation  and gets
radiated away through subsequent spontaneous decay to the ground state
(Shapiro  and Kang 1987), with the following cooling formula:
\begin{displaymath}
\Lambda_{H_2 \ Formation} = n_H (3.53 n_{H^-}k_8 + 1.38 n_{H_2^+})
        \ {\rm eV}\ \cm^{-3}\ \s^{-1}
\end{displaymath}
These reactions, however, are important heat sources at high densities
($n\geq 10^8 \ 1/cm^3$) where  collisional de-excitation can transform
most  of the excitation   energy  into kinetic (thermal)  energy.  The
density dependent  heating  rates are  given by  Hollenbach  and McKee
(1979, hereafter HM).

\paragraph{Line Cooling of Molecular Hydrogen}
\ 

Recently Martin, Schwarz, \& Mandy (1996, MSM96 herafter) have derived
the molecular hydrogen cooling function for \HH  - H collisions with a
complete set of rate coefficients.  Their result differs substantially
from former cooling  functions (e.g.  Hollenbach  and McKee 1979; Lepp
and Shull 1983). Especially  at number densities eceeding $10\cm^{-3}$
Lepp and Shull's expression seems to  overestimates (underestimate)
the cooling at low (high) temperatures by an order of magnitude.

Although, the   rate coefficients of    MSM96  are only  accurate  for
temperatures above $600\K$, some of them have been found to agree with
QCT calculations at   $300\K$.  For lower temperatures, however,   the
MSM96  cooling function has to be  understood as lower limit (personal
communication          with      Peter           Martin,     1996).    
\href{http://zeus.ncsa.uiuc.edu:8080/~abel/PGas/cool.html}{A FORTRAN77
  routine that computes  their fitting  formula}  can be found on  our
WWW-Site.  From Appendix  \ref{sec:OtPRatio}  we know that the  number
density of ortho-\HH n$_2^o$ can be very  small for applications where
our model is valid.  Therefore,  if one were  to estimate the  cooling
behavior to a  high  accuracy, a separate  treatment  of the  ortho and
para-\HH states would be required.

\paragraph{Photodissociation Heating}
\

Our  model includes the  photodissociation processes of \HH, processes
(27) and  (28). Black  and Dalgarno  (1977)  argue that,   for typical
radiation  fluxes,   the  photodissociation by    the two-step Solomon
process (27)  yields $0.4\eV$ per  atom pair.  Hence the corresponding
heating function is
\begin{eqnarray}
\Gamma_{27} = 6.4\tento{-13} \nd{\mHH} \k{27} \erg  \cm^{-3} \s^{-1}.
\end{eqnarray}

For direct dissociation  into  the continua  of  the Lyman and  Werner
systems, the  situation is similar  to photoionization  heating.  Since
the  reaction  channel leading   to  the  excitation of   the produced
hydrogen atoms is only accessible  for excess photon energies  greater
than $10.2\eV$, we  can safely assume that  all the excess energy  of the
dissociating photon will be shared as kinetic energies of the produced
hydrogen atom pair. Hence the heating functions can be written,
\begin{eqnarray}
\Gamma_{28} =  \nd{\mHH} \int^{\infty}_{\nu_{th}} 4\pi \sigma_{28}(\nu) 
   \frac{i(\nu)}{h\nu} (h\nu - h\nu_{th}) d\nu,
\end{eqnarray}
where $i(\nu)$ denotes the  radiation  intensity and the integral   is
evaluated  for the ortho and  para-\HH component separately because of
their different threshold energies.

\section{Application and Discussion of the Model}
In this section we will use data from a 1d high  resolution study of a
cosmological sheet   with wavelength 4Mpc, to  discuss non-equilibrium
effects and  the  chemical dynamics  in primordial gas.   The  data is
taken from Anninos \& Norman 1996.

\begin{figure}[hb]
\centerline{\psfig{file=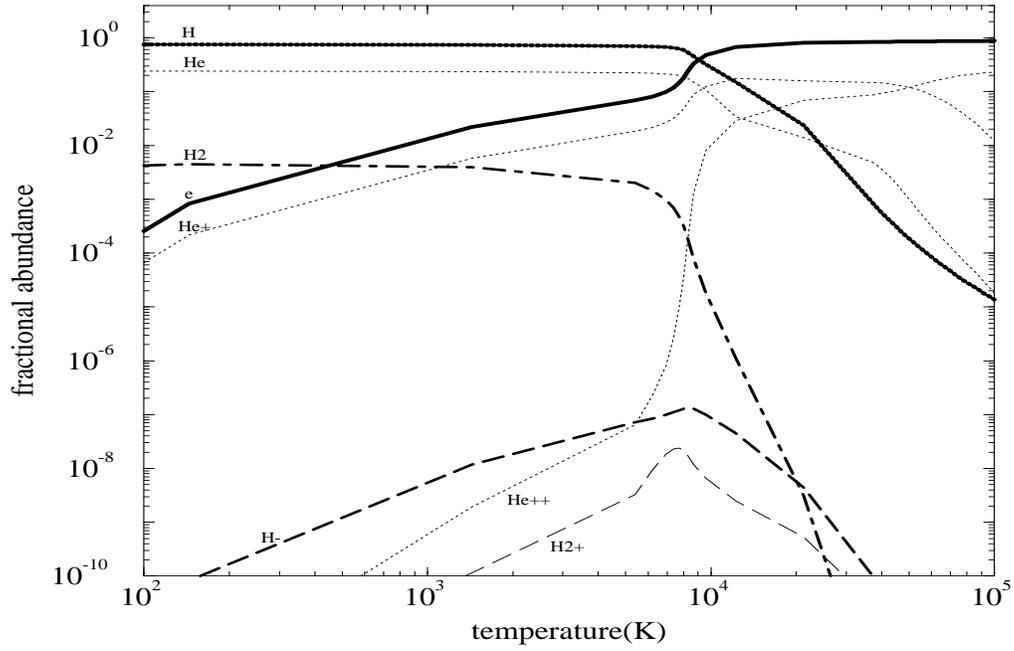,height=10cm,width=15cm}}
\caption{The fractional abundance of  nine species in a collapsing
  pancake of wavelength $4\Mpc$  are shown vs.\ temperature. The  data
  is taken from Anninos and Norman (1996). Clearly the non-equilibrium
  enhancement of free electrons at  low temperatures can be seen. Note
  also the similar non-equilibrium behavior of the \Hepp fraction.  }
\label{Fig:FracAbundances}
\end{figure}

Very  massive  pancakes have  strong  virialization shocks  ($v_s \sim
100$km/s),  which leads to  the   destruction of primordial  pre-shock
hydrogen   molecules.  The   post  shock  gas  cools   faster than  it
recombines and  leaves    the gas with  a   significant  free electron
fraction even at low temperatures.  
%
% figure \ref{Fig:FracAbundances} goes here
%
This can clearly be seen in Figure
\ref{Fig:FracAbundances}  where  the   fractional  abundances of   all
species  in the   post shock   gas  are  plotted   as a  function   of
temperature.  There  we  also see  that H$_2^+$   is always much  less
abundant than  H$^-$.  The doubly  negative helium ion recombines very
fast and therefore shows  the non-equilibrium effect at somewhat lower
temperatures than hydrogen.  These abundances agree well with the ones
given in Shapiro and Kang (1987).

\subsection{H$_2$ Chemistry}\label{sec:H2chem}

\begin{figure}[hb]
\centerline{\psfig{file=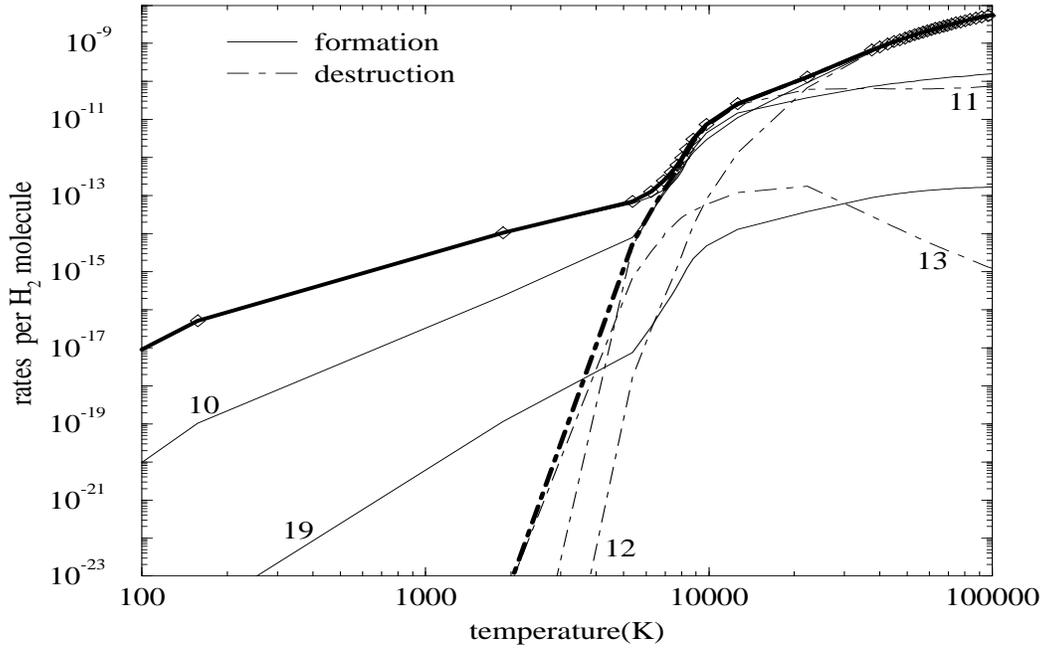,height=10cm,width=15cm}}
\caption{ The inverse of the reaction time scale for all
  processes in our   model, that involve    hydrogen molecules.  These
  relative  rates for the  producing collisional processes, (8), (10),
  and  (19),  are  illustrated by  solid  lines.   The \HH  destroying
  processes, (11),  (12), and (13),  are shown with  dot-dashed lines. 
  The  thick solid (thick dot-dashed)  line depicts the inverse of the
  sum   of  reaction time scales  of   all  \HH producing (destroying)
  processes. The divergence of  these producing and  destroying curves
  at low  temperatures  indicates that the  \HH  molecules  are out of
  equilibrium.  }\label{Fig:H2Chemistry}
\end{figure}

The right hand side of an individual rate equation is, for collisional
processes, given by a sum of terms $k_{ij} n_i n_j$, where $k_{ij}$ is
the rate  coefficient for  the reaction  between species  $i$ and $j$,
which  is   taken   positive  if   it  produces   the  species   under
consideration, or negative  if it destroys it.   To see what processes
dominate the   chemistry of  species $l$,   let us  consider  the rate
$k_{ij}  n_i n_j$  divided by $n_l$,  which  we call rate per $l$-atom
(molecule).     Comparing   this   to     the  sum    over     all $l$
producing/destroying  processes (the evaluated  right  hand side) much
insight about  the  ongoing processes  can  be  gained.  Actually this
quantity equals  the inverse reaction  time scale and  hence provides a
measure of  the   time  in which  a   particular  species  gets   into
equilibrium.   We  show   such plots  for   \HH,  \HHp, and  \Hm (Fig.
\ref{Fig:H2Chemistry},            \ref{Fig:H2+Chemistry},          and
\ref{Fig:HMChemistry}).
%
% figure \ref{Fig:H2Chemistry} goes here
%
 The complete H$_2$ chemistry  in  our model  is illustrated in   Fig.
\ref{Fig:H2Chemistry}.  All molecular hydrogen producing reactions (8,
10, and 19)  are shown as solid  lines as well  as  their sum (labeled
formation).  The destroying processes (11, 12, and 13) are depicted by
dashed --  dotted lines.   It   is clear that  the  molecular hydrogen
abundance at  temperatures  above $\sim  7000\K$   is a result  of  the
balance between producing   and destroying processes,   what one could
call a  ``collisional equilibrium''.  It   can also be  seen that from
$0.5$eV   to $1$eV  the charge  exchange  reaction  of H$_2$ and  free
protons is the most efficient molecular hydrogen destruction path.  At
higher temperatures, molecular hydrogen are destroyed most efficiently
by free electrons. Obviously here in the pancake collapse, destruction
by neutral hydrogen atoms has  a negligible influence on the molecular
hydrogen   abundance, which    is     certainly due  to      the  high
(non-equilibrium) abundance of free protons which will not be found as
drastically enhanced, compared to the equilibrium abundances, for weak
shock waves.  For  the production of H$_2$  we find that proc.   8 and
19, are always dominated by the fast H$^-$ formation path (proc. 8).

In  summary, we find that the  molecular hydrogen fraction is governed
dominantly  by three  processes.    Formation via  H$^-$,  destruction
through  charge  exchange with free  protons,  and destruction by free
electrons.   These insights can be used  also for analytical estimates
of the molecular hydrogen  fraction  during pancake collapse.  Let  us
illustrate this   briefly on  the  molecular hydrogen    abundance for
temperatures, $7000\K\leq T \leq 10^4\K$. Here the H$_2$ abundance
can     obviously  be derived   through     the equilibrium  condition
$dn_{H_2}/dt = 0$ with,
\begin{eqnarray*}
\frac{dn_{H_2}}{dt} \simeq k_8 n_{H^-} n_H - k_{11} n_{H_2} n_{H^+} = 0,\ 
  \Rightarrow \  n_{H_2} \simeq \frac{k_8}{k_{11}} \frac{n_{H^-}}{x}
\end{eqnarray*}
where  $x$ denotes  the    ionized fraction.   Looking   up   the rate
coefficients we can read  off  the qualitative temperature  dependence
and find that with a decreasing temperature in  the range $6000\K \leq
T \leq 2\times 10^4\K$,   the molecular hydrogen  abundance increases.
For lower temperatures, collisional processes are not efficient enough
in destroying  molecular hydrogen and it will  get produced as long as
there exists a significant amount of H$^-$ ions at the  level of a few
percent.

\subsection{The \Hm and \HHp Chemistry}

%
% figure \ref{Fig:HMChemistry} goes here
%

\begin{figure}[hb]
\centerline{\psfig{file=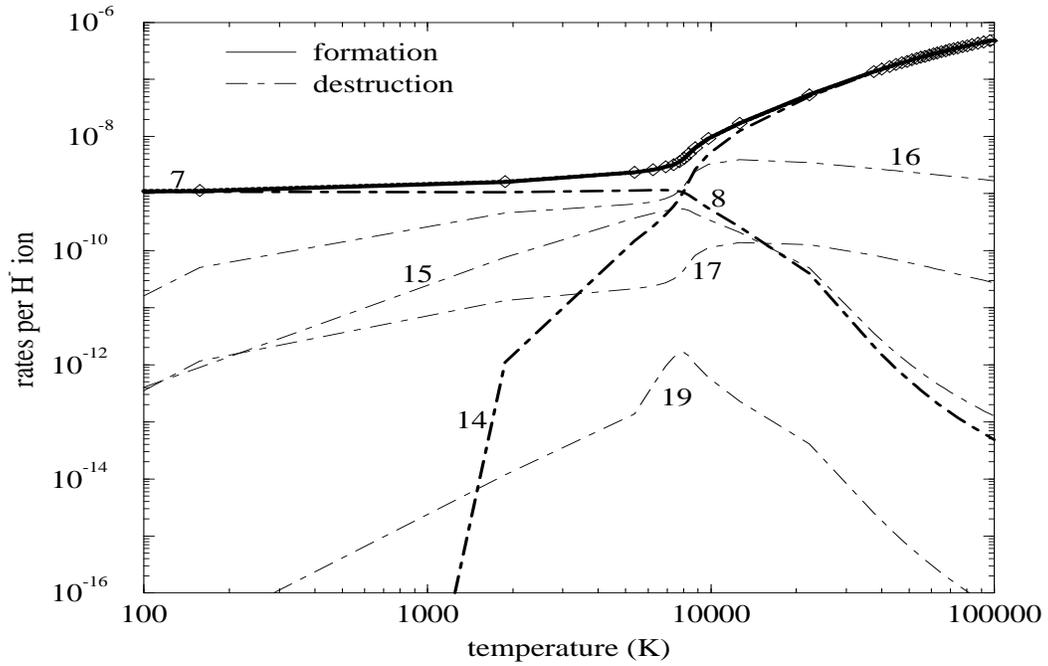,height=10cm,width=15cm}}
\caption{ The relative rates
  are shown for the  collisional processes involving  H$^-$, including
  the  single  \Hm producing  reaction  (7), and  all  its destruction
  processes (8), (14), (15), (16), (17), and (19).  It is evident that
  \Hm reached its equilibrium  abundance within the hydrodynamical and
  cooling time scale since the sum  of the production rates equals the
  sum of the destruction  rates.  Here the  thick dot-dashed lines
  illustrate the major  \Hm    destroying processes,  (8)  and (14).   
  }\label{Fig:HMChemistry}
\end{figure}

\begin{figure}[hb]
\centerline{\psfig{file=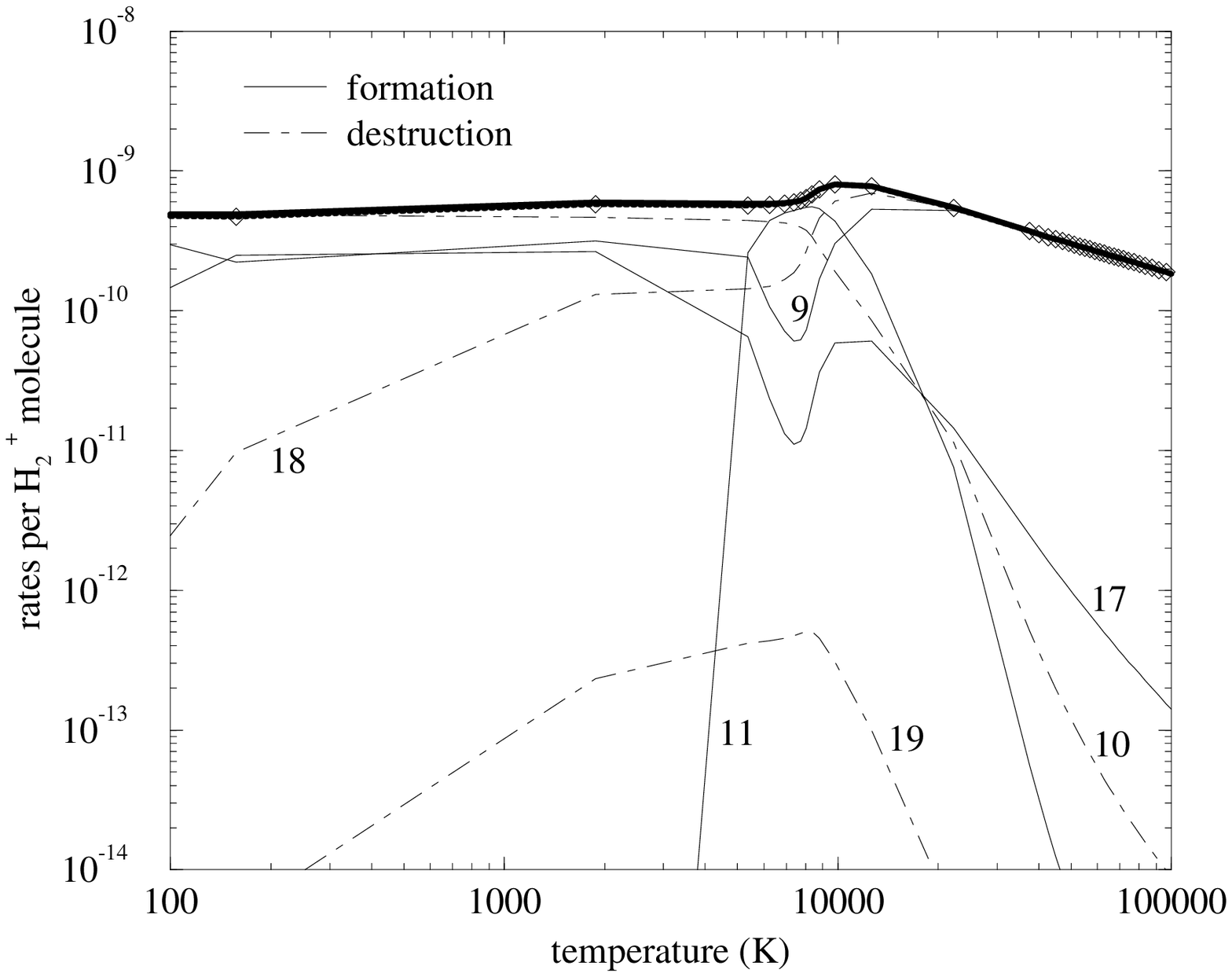,height=10cm,width=15cm}}
\caption{ The inverse  of the reaction time scale for the
  collisional processes  involving \HHp.  The \HHp producing processes
  (9), (11),  and (17)  are  depicted  by solid  lines  and  the  \HHp
  destroying reactions  (10,  (18),   and   (19) correspond   to   the
  dot-dashed lines.  Analogous  to \Hm (figure ~\ref{Fig:HMChemistry}),
  it  is clear that \HHp reached  its equilibrium abundance within the
  hydrodynamical and cooling time scales. }\label{Fig:H2+Chemistry}
\end{figure}

The H$^-$ ion  is found to get into  chemical  equilibrium faster than
the dynamical time  scales    (Courant, free-fall,  and   cosmological
times).  Hence  the overall destruction and production  lines  in Fig. 
\ref{Fig:HMChemistry} are right   on top of   each  other.  Our  model
includes one H$^-$ producing reaction, the radiative attachment (proc.
7), and six destroying reactions.  Fig.  \ref{Fig:HMChemistry} clearly
illustrates that the  processes 17 and  19 play no  role, processes 15
and 16, little role, and processes 7, 8, and 14 the main role, for the
H$^-$ chemistry.  Since the atomic data we used for the main processes
7, 8, and 14 are very accurate at all temperatures we ensured that the
model will give also accurate predictions for the H$^-$ abundance.
%
% figure \ref{Fig:H2+Chemistry} goes here
%
An analogous  plot  for   the  \HHp chemistry  is   given  with  Fig.  
\ref{Fig:H2+Chemistry}.   Obviously    \HHp   has  been   in  chemical
equilibrium  for  all temperatures in  our pancake  study.  All of its
three production   mechanisms contribute   in  individual  temperature
regimes.  The  radiative  association  reaction proc.   9  contributes
strongly for temperatures below a few thousand Kelvin and dominates at
higher  temperatures as $\sim 2   \times 10^4\K$.   For the  remaining
interval  charge  exchange between  H$_2$   and H$^+$ is the  dominant
H$_2^+$ producing reaction.     Furthermore  it is  clear  from   Fig. 
\ref{Fig:H2+Chemistry}  that  one can  safely  neglect  the destroying
mechanism proc. 19, because it  does not contribute significantly,  at
any  temperature, to the  overall destruction of  H$_2^+$.  It is also
evident  that at all temperatures  below $\sim 8 \times 10^3\K$ nearly
every  H$_2^+$ molecule that  gets  produced  will be  converted  to a
neutral hydrogen molecule by proc. 10.  At higher temperatures H$_2^+$
will be destroyed by electrons.   The data of  Schneider \etal for the
rate  coefficient of this process, and  consequently our fit, is quite
inaccurate at such high temperatures and we  do not expect the H$_2^+$
abundance at temperatures higher than  $10^4$K to be reliable.  We see
that process  19 was never  important in determining the abundances of
H$^-$, H$_2^+$, and of H$_2$, and can therefore be neglected in future
applications.

\subsection{A Minimal Model}

From the preceding  section we clearly see  that not all reactions are
equally important.  We devised a  minimal model that incorporates only
the  essential processes need  to   accurately model the formation  of
molecular hydrogen.   From Figure \ref{Fig:H2Chemistry} it  is evident
that  the only reaction involving  \HHp which  contributes strongly to
the \HH chemistry is  the charge exchange  reaction (11).  Its rate is
independent of  the   \HHp abundance since   it  is a \HHp   producing
reaction and we  can conclude that at  least for strong shocks the \HH
abundance  is independent of the \HHp  chemistry.  For a minimal model
we can use this  and leave out  reactions 9, 10, 17,  and 19 so as  to
avoid solving  a  \HHp rate equation.  Further   by looking at  Figure
\ref{Fig:HMChemistry} we find the  process  (15) to be  negligible and
process (16) only to be marginally important. Clearly (13) can be left
out as    well since it  obviously does   not  contribute to   the \HH
chemistry.  Furthermore, we note  that it is  not necessary to include
the contributions from the reaction involving \Hm and \HH to the major
species H,   He,  and  their  ions.    With these simplifications   we
eliminated  7 of the 19 reactions  and reduced the reaction network to
the following,

\begin{eqnarray}
\frac{d \nd{H} }{dt}    &=& \k{2} \nd{\mHp} \nd{e} - \k{1} \nd{H}    \nd{e}     \\
\frac{d\nd{\mHp}}{dt}  &=& \k{1} \nd{H}   \nd{e} - \k{2} \nd{\mHp}  \nd{e}      \\
\frac{d\nd{He} }{dt}  &=& \k{4} \nd{\mHep}\nd{e} - \k{3} \nd{He}   \nd{e}       \\
\frac{d\nd{\mHep}}{dt} &=& \k{3} \nd{He} \nd{e} + \k{6} \nd{\mHepp}\nd{e} -
                       \k{4} \nd{\mHep}\nd{e}                           \\
\frac{d\nd{\mHepp}}{dt}&=& \k{5} \nd{\mHep}\nd{e} - \k{6} \nd{\mHepp} \nd{e}    \\
\frac{d\nd{\mHH} }{dt}  &=& \k{8} \nd{\mHm} \nd{H} - \nd{\mHH}   \left( 
        \k{11} \nd{\mHp} + \k{12} \nd{e} \right) , 
\end{eqnarray}
where the number density of \Hm is given by the equilibrium condition
\begin{eqnarray}
\nd{\mHm} = {\k{7} \nd{H} \nd{e}  \over \k{8} \nd{H}  + 
 \k{16} \nd{\mHp} + \k{14} \nd{e}}.
\end{eqnarray}
To  derive the  number   denisty of  \Hm by its   equilibrium value is
justified   by  comparing   the   figures, \ref{Fig:H2Chemistry}   and
\ref{Fig:HMChemistry}  .  Since they show  the inverse of the reaction
time scales of the different  processes it is  clearly evident that all
reactions      determining the \Hm    abundance  occur  on much faster
time scales  than  those  responsiblefor  the \HH   chemistry.  We have
checked this  minimal model extensively and  find the \HH abundance to
generally aggree with the full model to  within a few percent over the
entire temperature of the individaul application.

\section{Conclusions}

We have derived a reliable  time dependent chemistry and cooling model
for  primordial gas that, in connection  with our new numerical method
(AZAN96),    proves  as a   powerful  tool  to  investigate primordial
structure  formation in multidimensional  numerical calculations.  The
model is designed to  be applicable for densities below $10^4\cm^{-3}$
and  temperatures $<10^8\K$.  We have  discussed  the influence of the
orth-\HH to   para-\HH  ratio on the   cooling behaviour  of  the gas,
derived new fits to molecular   data, and  developed a minimal   model
capable of describing primordial gas in applications where no external
radiation fields are considered.

\section{Related Websites}
At our WWW-Site,
\href{http://zeus.ncsa.uiuc.edu:8080/~abel/PGas/LCA-CM.html}{The LCA
Cooling Model}, we present all the used atomic and molecular data in
great detail. Further we provide there FORTRAN routines that compute
the rate coefficients and sove the rate equations.  Tom Abel continues
to collect molecular data related to primordial gas at 
\href{http://zeus.ncsa.uciu.edu:8080/~abel/PGas/}{The Primordial Gas
Chemistry Page}.  Further
\href{http://www.pa.uky.edu/~verner/atom.html}{ Dima Verners Atomic
Data Page} represents a superb reference for atomic data.

\acknowledgments

We would  like to thank Dimitri  Mihalas, Mordecai-Mark MacLow, Zoltan
Haiman,  Max  Tegmark, and  Evelyne   Roueff  for helpful  discussion.
Furthermore,   we are grateful    for useful correspondence with  Bill
Langer, Phil Stancil, Jonathan  Rawlings, Alex Dalgarno, Peter Martin,
and Stephen Lepp.  This work was done partly under the auspices of the
Grand Challenge Cosmology Consortium funded  by NSF grant ASC-9318185.
Tom  Abel happily   acknowledges the  hospitality  of  the Max  Planck
Institut f\"ur Astrophysik and the   Insituto Astrofisica de  Canarias
where parts of this work have been completed.

\clearpage
\appendix
\section{Reactions and Rates}\label{app:rates}

Here we present all the rate coefficients  used in our model. All fits
are accurate  for temperatures ranging from 1K  to $10^8\K$.  Since we
are interested in numerical applications we  did pay more attention to
the  accuracy  of the  fits than to  the   simplicity of the formulas.
\href{http://zeus.ncsa.uiuc.edu:8080/~abel/PGas/LCA-CM.html}{A FORTRAN
program that computes these rate coefficients} can  be obtained at our
WWW site. The temperatures are in $\eV$ unless stated otherwise.

{\scriptsize
%{\small
\begin{table}[ht]
\caption{\label{tab:rates}
{\sl Collisional Ionization and Radiative Recombination of Hydrogen
  and Helium. }}
 
(1) \ \ \underline{H  \ + \  e$^-$ \   $\rightarrow$ \   H$^+$ \  + \  2e$^-$} \hfill  Janev et al. 1987 (2.1.5) \hspace{1.0cm}
\begin{eqnarray*}       
 k_1 & = & \exp[-32.71396786 + 13.536556  \ln(T) - 5.73932875   \ln(T)^2 + 1.56315498  \ln(T)^3 -       \\
   &  & 0.2877056   \ln(T)^4 + 3.48255977\times 10^{-2} \times\ln(T)^5- 2.63197617\times 10^{-3}\times\ln(T)^6 +   \\
   & &  1.11954395\times 10^{-4}   \ln(T)^7 -  2.03914985 \times 10^{-6}   \ln(T)^8] \ \cm^3 \s^{-1}.
\end{eqnarray*}
(2)  \ \  \underline{H$^+$ \  + \  e$^-$ \  $\rightarrow$ \  H \  + \  $\gamma$} \hfill Our fit to data given by Ferland et al. (1992) \hspace{1.0cm}    
\begin{eqnarray*}
 k_2 &=& \exp[-28.6130338 - 0.72411256  \ln(T) - 2.02604473\times10^{-2}  \ln(T)^2 - \\
  & &    2.38086188\times10^{-3}  \ln(T)^3 - 3.21260521\times10^{-4}  \ln(T)^4 - 1.42150291\times10^{-5}  \ln(T)^5 + \\
   & &  4.98910892  \times10^{-6}   \ln(T)^6 + 5.75561414 \times10^{-7}   \ln(T)^7 - 1.85676704 \times10^{-8}   \ln(T)^8 -\\
  & &   3.07113524 \times10^{-9}   \ln(T)^9] \ \cm^3 \s^{-1}.
\end{eqnarray*}
(3)  \ \ \underline{He     \  + \ \   e$^-$\ \ $\rightarrow$ \ He$^+   $\ \ + \ \  2e$^-$}  \ \hfill  Janev et al. 1987 (2.3.9) \hspace{1.0cm}    
\begin{eqnarray*}
 k_3 &=& \exp[ (-44.09864886 + 23.91596563   \ln(T) -  10.7532302   \ln(T)^2 + 3.05803875   \ln(T)^3 -\\
   & & 0.56851189   \ln(T)^4 + 6.79539123 \times 10^{-2}  \ln(T)^5 - 5.00905610 \times 10^{-3}   \ln(T)^6 + \\
   & & 2.06723616 \times 10^{-4}   \ln(T)^7 - 3.64916141 \times 10^{-6}   \ln(T)^8) \ \cm^3 \s^{-1}.
\end{eqnarray*}
(4)  \ \ \underline{He$^+$\ \ + \ \   e$^-$\ \ $ \rightarrow$\ He     \ \ + \ \  $\gamma$}  \ \hfill  Cen (1992) and Aldrovandi \& Pequignot (1973) \hspace{1.0cm}    
\begin{eqnarray*}
{\rm  Radiative: \ \ \ \ } k_{4r} & = & 3.925 \times 10^{-13}T^{-0.6353}\ \cm^3 \s^{-1} \\
{\rm Dielectronic: \ }k_{4d} &=& 1.544 \times 10^{-9} T^{-\frac{3}{2}} 
     \exp\left( - \frac{48.596\eV}{T} \right) \
        \left[0.3 + \exp\left(\frac{8.10\eV}{T}\right)\right] 
   \ \cm^3 \s^{-1}.
\end{eqnarray*}
(5)  \ \  \underline{He$^+   $\ \ + \ \   e$^-$\ \ $ \rightarrow$\ He$^{++}$\ \ + \ \  2e$^-$\ } \hfill  Aladdin Database (1989) \hspace{1.0cm}    
\begin{eqnarray*}
k_5 & = & \exp[-68.71040990 + 43.93347633   \ln(T) - 18.4806699   \ln(T)^2 + 4.70162649   \ln(T)^3 - \\
 & &    0.76924663   \ln(T)^4 + 8.113042 \times 10^{-2}  \ln(T)^5 - 5.32402063 \times 10^{-3}  \ln(T)^6 + \\
 & &    1.97570531\times 10^{-4}  \ln(T)^7 -  3.16558106 \times 10^{-6}  \ln(T)^8] \ \cm^3 \s^{-1}.
\end{eqnarray*}
(6)  \ \  \underline{He$^{++}$\ \ + \ \   e$^-$\ \ $ \rightarrow$\ He$^+   $\ \ + \ \ $\gamma$} \hfill Cen (1992) \hspace{1.0cm}  
\begin{eqnarray*}
k_6 = 3.36 \times 10^{-10} T^{-\frac{1}{2}}
\left(\frac{T}{1000\K}\right)^{-0.2}
\left(1+\left(\frac{T}{10^6\K}\right)^{0.7}\right)^{-1} \ \cm^3
\s^{-1},  {\rm \ \ where\ T\ is\ in\ K.}
\end{eqnarray*}

\end{table}

\clearpage
\begin{table}[ht]
\caption{\label{tab:rates2} 
{\sl  The formation paths of {\rm H}$_2$.}}

(7)  \ \  \underline{H \ \ + \ \  e$^-  $\ \ $ \rightarrow$\ H$^-$\ \ + \ \ $\gamma$} \ \hfill  This work from data by Wishart (1979)  \hspace{1.0cm}   
\begin{eqnarray*} 
\ k_7 = \left\{ \begin{array}{ll}
           1.429 \times 10^{-18} T^{0.7620} T^{0.1523 \log_{10}(T)}  T^{-3.274 \times 10^{-2} \log_{10}^2(T)}  \ \cm^3 \s^{-1} 
                  {\rm \ \ \ for \ \ } T \leq 6000 \K  \\
           3.802 \times 10^{-17} T^{0.1998 \log_{10}(T)} {\rm dex}\left({4.0415 \times 10^{-5} \log_{10}^6(T)  - 5.447 \times 10^{-3} \log_{10}^4(T)}\right)  \ \cm^3 \s^{-1} 
                 &  {\rm \ \ otherwise.}
              \end{array}
      \right.
 \end{eqnarray*}
(8)  \ \  \underline{H \ \ +  \   H$^-$\ \ $\rightarrow$\  H$_2 $\ \ + \ \  e$^-$}   \ \hfill  Our Integration of data from Janev et al. (1987) \hspace{1.0cm} 
\begin{eqnarray*}
T&>&0.1\eV:\ k_8 = \exp[-20.06913897 +0.22898   \ln(T) +  3.5998377\times 10^{-2}   \ln(T)^2 - \\
 && 4.55512 \times 10^{-3}  \ln(T)^3- 3.10511544\times 10^{-4}   \ln(T)^4  + 1.0732940 \times 10^{-4}  \ln(T)^5 -   \\
 &&  8.36671960 \times 10^{-6}   \ln(T)^6 + 2.23830623 \times 10^{-7}   \ln(T)^7]  \ \cm^3 \s^{-1}. \\
T &<& 0.1\eV:  k_8  =  1.428 \times 10^{-9} \ \cm^3 \s^{-1}.
\end{eqnarray*}
(9)  \ \  \underline{H \ \ \ + \ \ H$^+$ \ $\rightarrow$\ H$_2^+$\ \ +\ \ $\gamma$}  \hfill Shapiro \& Kang (1987) \hspace{1.0cm}  
\begin{eqnarray*}
\ \ k_9 = \left\{ \begin{array}{ll}
        3.833 \times 10^{-16} \times T^{1.8} \ \cm^3 \s^{-1},
                                            &  {\rm for \ \ } T< 0.577 \eV, \\
        5.81 \times 10^{-16} \times (0.20651 \ T)^{-0.2891 \times 
                \log(0.20651 \times T)} \ \cm^3 \s^{-1},
                                            &  {\rm for \ \ } T \geq 0.577 \eV.
              \end{array}
      \right.
\end{eqnarray*}
(10) \  \underline{H$_2^+$\ \ + \ \ H \ \ $\rightarrow$\ H$_2$\ \ + \ \ H$^+$} \hfill  Karpas et al. (1979)     \hspace{1.0cm}                          
\begin{eqnarray*}\ \ k_{10} = (6.4 \pm 1.2) \times 10^{-10} \ \cm^3 \s^{-1}.\end{eqnarray*}
\end{table}

\clearpage
\begin{table}[ht]
\caption{\label{tab:rates3}
{\sl Other Collisional Processes involving {\rm H}$^-$, {\rm H}$_2^+$, and {\rm H}$_2$.}}

(11) \  \underline{H$_2$\ \ + \ \ H$^+$\ \ $\rightarrow$ \ H$_2^+$\ \ + \ \ H} \hfill  This work \hspace{1.0cm}    
\begin{eqnarray*}
\ln(k_{11})& &=    -24.24914687    +
        3.40082444  \ln(T)    -
        3.89800396  \ln(T)^2    +   
        2.04558782  \ln(T)^3    - \\
   & &     0.541618285  \ln(T)^4   + 
        8.41077503 \times 10^{-2}  \ln(T)^5  -
        7.87902615 \times 10^{-3} \ln(T)^6   + \\
   & &      4.13839842 \times 10^{-4} \ln(T)^7   - 
       9.36345888\times 10^{-6}  \ln(T)^8 \ \cm^3 \s^{-1}.
\end{eqnarray*}
(12) \ \underline{H$_2$\ \ + \ \ e$^-$\ \ $\rightarrow$ \ 2H \ \ + \ \
e$^-$} \hfill Donahue and Shull (1991) \hspace{1.0cm}
\begin{eqnarray*}
\ \ k_{12} = 5.6 \times 10^{-11} T^\frac{1}{2} \exp(-\frac{102,124\K
}{T}) \ \cm^3 \s^{-1}, {\rm \ \ where\ T\ is\ in\ K.}
\end{eqnarray*}
(13) \  \underline{H$_2    $\ \ + \ \ H\ \ $\rightarrow$  \ 3H}   \hfill  Dove and Mandy (1986) \hspace{1.0cm}    
\begin{eqnarray*}\ \ k_{13} = 1.067 \times 10^{-10}  T^{2.012}\times \exp (-(4.463/T) (1 + 0.2472 \ T )^{3.512} ) \ \cm^3 \s^{-1}.
\end{eqnarray*}         
(14) \  \underline{H$^-$\ \ +\ \ e$^-$\ \ $ \rightarrow$ \ H\ \ + \ \ 2 $e^-$} \hfill  Janev et al. (1987, 7.1.1)       \hspace{1.0cm}          
\begin{eqnarray*}
k_{14}&=&\exp[-18.01849334 + 2.3608522   \ln(T) - 0.28274430   \ln(T)^2 + 1.62331664 \times 10^{-2}  \ln(T)^3 -  \\
  &&   3.36501203 \times 10^{-2}   \ln(T)^4 + 1.17832978 \times 10^{-2}   \ln(T)^5 - 1.65619470 \times 10^{-3}   \ln(T)^6 + \\
  & &   1.06827520 \times 10^{-4}   \ln(T)^7 - 2.63128581 \times 10^{-6}   \ln(T)^8) \ \cm^3 \s^{-1}.
\end{eqnarray*}
(15) \  \underline{H$^-$\ \ + \ \ H\ \ $\rightarrow$ \ \ 2H \ \ + \ \  e$^-  $} \hfill  Our modification to the Janev et al. (1987) data\hspace{1.0cm}
\begin{eqnarray*}
T>0.1\eV&:&   k_{15} =\exp[-20.37260896 + 1.13944933  \ln(T) -  0.14210135  \ln(T)^2 +  \\
  & &   8.4644554 \times 10^{-3} \ln(T)^3 - 1.4327641\times 10^{-3}  \ln(T)^4 + 2.0122503 \times 10^{-4} \ln(T)^5 +  \\
  & &   8.6639632 \times 10^{-5} \ln(T)^6 - 2.5850097\times 10^{-5} \ln(T)^7 + 2.4555012 \times 10^{-6}  \ln(T)^8 - \\
  & &   8.0683825 \times 10^{-8}  \ln(T)^9] \ \cm^3 \s^{-1}, \\
 T<0.1\eV&:&   k_{15} = 2.5634 \times 10^{-9} \times T^{1.78186} \ \cm^3 \s^{-1}.
\end{eqnarray*}
(16) \  \underline{H$^-$\ \ + \ \ H$^+$\ \ $\rightarrow$ \ 2H } \hfill  Dalgarno and Lepp (1987) \hspace{1.0cm}                         
\begin{eqnarray*}\ \ k_{16} = 7 \times 10^{-8} \left( \frac{T}{100{\rm
        K}} \right)^{-\frac{1}{2}} \ \cm^3 \s^{-1}, {\rm \ \ where\ T\ is\ in\ K.}
\end{eqnarray*}
(17) \ \underline{H$^-$\ \ + \ \ H$^+$\ \ $\rightarrow$ \ H$_2^+$\ \ + \ \ e$^-$} \hfill  Our modification to fit of Shapiro and Kang (1987) \hspace{1.0cm}   
\begin{eqnarray*}
\ \ k_{17} = \left\{   \begin{array}{ll}
                  2.291\times 10^{-10}T^{-0.4} \ \cm^3 \s^{-1}, &{\rm for \ \ } T<1.719 \eV \\
                  8.4258\times 10^{-10} T^{-1.4} \exp(-1.301/T)\ \cm^3 \s^{-1}, 
                                        &{\rm otherwise. }
                \end{array}
        \right.
\end{eqnarray*}
(18) \ \underline{H$_2^+$\ \ + \ \ e$^-$\ \ $\rightarrow$\  2H} \hfill Our fit to the Schneider et al. (1994) data \hspace{1.0cm}    
\begin{eqnarray*}
\ \ k_{18} = \left\{ \begin{array}{lll}
     1.0 \times 10^{-8} \ \cm^3 \s^{-1}, 
        &   {\rm for \ }  & T < 617 \K , \\
     1.32 \times 10^{-6} T^{-0.76} \ \cm^3 \s^{-1}, 
        &   {\rm for \ \ }& T > 617 \K,  {\rm \ \ where\ T\ is\ in\ K.}
                \end{array}
        \right.
\end{eqnarray*}
(19) \ \underline{H$_2^+$\ \ + \ \ H$^-$\ \ $\rightarrow$\ H$_2$\ \ + \ \ H} \hfill  Dalgarno and Lepp (1987) \hspace{1.0cm}    
\begin{eqnarray*}
\  \  k_{19}  = 5  \times  10^{-7}   \left(  \frac{100\K}{T}  \right)^{
\frac{1}{2} }\ \cm^3 \s^{-1},  {\rm \ \ where\ T\ is\ in\ K.}
\end{eqnarray*}

\end{table}

\clearpage
\begin{table}[ht]
\caption{{\sl Photoionization and Photodissociation.}}\label{tab:rates4}

(20) \ \   \underline{H \ \ \ + \ \ $\gamma$\ \ $\rightarrow$\  H$^+$ + \ \ e$^-$},\ 
\\ 
(22) \ \  \underline{He$^+ $\ \ $+ \ \ \gamma $\ \ $\rightarrow$\ He$^{++}$\ \ + \ \ e$^-$}  \hfill  Osterbrock (1974)  \hspace{1.0cm}   
\begin{eqnarray*}
\sigma_{20,22} = \frac{A_0}{Z^2} \left( \frac{\nu}{\nu_{th}} \right)^4  \frac{\exp[4-4(\arctan \epsilon)/\epsilon]}{1-\exp(-2\pi/\epsilon)}, 
        {\rm where \ } A&=&6.30 \times 10^{-18} \cm^2,\\ \epsilon&=&\sqrt{\nu/\nu_{th}-1},\ h\nu_{th}=13.6 Z^2 \eV.
\end{eqnarray*}
(21) \ \  \underline{He \ \ \ + \ \ $\gamma$\ \ $\rightarrow$\ He$^+ $\ \ + \ \ e$^- $}       \hfill Osterbrock 1974 \hspace{1.0cm} 
\begin{eqnarray*}
\sigma_{21}^{\nu} = 7.42 \times 10^{-18} \left( 1.66 \left(
        \frac{\nu}{\nu_{th}} \right)^{-2.05}  -  0.66 \left(
        \frac{\nu}{\nu_{th}} \right)^{-3.05}  \right)\cm^2, {\rm \ for \ } \nu>\nu_{th}.
\end{eqnarray*}                                                                 
(23) \ \  \underline{H$^-$\ \ \ + \ \ $\gamma $\ \ $\rightarrow$\ H \ \ + \ \ e$^- $}   \hfill Shapiro and Kang (1987)\hspace{1.0cm}   
\begin{eqnarray*}
\sigma_{23} = 7.928 \times 10^5 (\nu - \nu_{th})^\frac{3}{2} \frac{1}{\nu^3} \cm^2, {\rm \ \ for \ } h\nu>h\nu_{th}=0.755 \ \eV.
\end{eqnarray*}
(24) \ \  \underline{H$_2 $\ \ \ + \ \ $\gamma $\ \ $\rightarrow$\ H$_2^+ $\ \ + \ \ e$^- $}  \hfill O'Neil and Reinhardt (1978) \hspace{1.0cm}   
\begin{eqnarray*}
\sigma_{24}^\nu = \left\{ \begin{array}{lll}
     0,                                             
        &   {\rm for \ }  & h\nu  < 15.42\eV , \\
     6.2 \times 10^{-18} h\nu - 9.4 \times10^{-17}\cm^2, 
        &   {\rm for \ \ }  15.42\eV \leq  & h\nu  < 16.50\eV, \\
     1.4 \times10^{-18} h\nu - 1.48 \times10^{-17}\cm^2, 
        &   {\rm for \ \ }  16.50\eV  \leq  & h\nu  < 17.7\eV, \\
     2.5 \times 10^{-14} (h\nu)^{-2.71} \cm^2 ,             
        &   {\rm for \ }  & h\nu  \geq 17.7\eV.
                \end{array}
        \right.
\end{eqnarray*}
(25) \ \  \underline{H$_2^+$\ \ \ + \ \ $\gamma $\ \ $\rightarrow$\ H \ \ +  \ H$^+ $}       \hfill Our fit to Stancil (1994)   \hspace{1.0cm}   
\begin{eqnarray*}
\log_{10}(\sigma_{25}^{GS}) & = &              
( -1.6547717\times 10^6  + 1.8660333\times 10^5 \ln(\nu) -
   7.8986431\times 10^3 \ln(\nu)^2 + \\
& & 148.73693  \ln(\nu)^3  -   1.0513032 \ln(\nu)^4 )
\end{eqnarray*}
(26) \ \  \underline{H$_2^+$\ \ \ + \ \ $\gamma $\ \ $\rightarrow$\ 2H$^+$\ \ +\  \ e$^-$}        \hfill Shapiro and Kang (1987) \hspace{1.0cm}   
\begin{eqnarray*}
\log_{10}(\sigma_{26})  &  =& -16.926 - 4.528    \times 10^{-2} h\nu +
2.238 \times 10^{-4}  (h\nu)^2 + 4.245 \times  10^{-7} (h\nu)^3,\\ & &
{\rm \ \ \ \ \ \ for \ }30\eV<h\nu<90\eV.
\end{eqnarray*}
(27) \ \  \underline{H$_2 $\ \ \ + \ \ $\gamma $\ \ $\rightarrow$\ H$_2^*$\  $\rightarrow$\  H \ \ +  \ H}   \hfill This work  \hspace{1.0cm}   
\begin{eqnarray*} 
{\rm  Neglecting\  selfshielding,} & k_{27} = 1.1   \times  10^8
 j(\overline{\nu})\  \s^{-1},       \\   {\rm     \  \       where    \
 }j(\overline{\nu}),{\rm\  is\ the\ radiationflux}   &{  \rm in\
 ergs \ s}^{-1} \ \cm^{-2}{\rm \ at\ h\overline{\nu}=12.87\eV }.
\end{eqnarray*} 
(28) \ \  \underline{H$_2 $\ \ \ + \ \ $\gamma $\ \ $\rightarrow$\ H \ \ +  \ H}              \hfill This work  \hspace{1.0cm}   
\begin{eqnarray*} 
\sigma_{28} = \frac{1}{y+1} (\sigma_{28}^{L0} + \sigma_{28}^{W0}) + 
   (1-\frac{1}{y+1}) (\sigma_{28}^{L1} + \sigma_{28}^{W1})
\end{eqnarray*} 
\end{table}
}

\clearpage
\section{Ortho-\HH to para-\HH Ratio} \label{sec:OtPRatio} 

Many of the  collisional rates have  only  been computed for  the para
configuration and  are applicable only  if para-\HH is is the dominant
modification (Flower and Watt 1984, hereafter FW). For our purposes it
plays another significant role   in  that its selection   criteria for
allowed rotational transitions  are  different.   This is  crucial  to
determine the exact cooling  behavior  at low temperatures,  since low
density primordial gas will cool mostly in  the rotational lines $E_3$
--  $E_1$  $=844.5$K for the   ortho configuration and $E_2$  -- $E_0$
$=509.9$K for para-\HH.  At  formation the two different modifications
will be abundant corresponding to their statistical weights, namely
\begin{eqnarray}
\frac{n(ortho)}{n(para)} = \frac{n(J=1)}{n(J=0)} = 3.
\end{eqnarray}
FW argue for a single interconversion mechanism at low densities,
\begin{eqnarray}\label{equ:oTOpR}
\rm H_2({\rm ortho}) \ \ + \ \ H^+ \ \ \rightarrow \ \ \rm H_2({\rm
para}) \ \ + \ \ H^+ \ \ + \ \ 170.5 \K,
\end{eqnarray}
and suggest a temperature independent rate coefficient of, $k_{op} = 3
\times   10^{-10}  \cm^3   \s^{-1}$.  If   the  $J=2$  state  would be
significantly populated, para-\HH could be transformed to ortho by,
\begin{eqnarray*}
\rm H_2(J=2) \ \ + \ \ H^+ \ \ \rightarrow \ \ \rm H_2(J=1) \ \ + \ \
H^+ \ \ + \ \ 170.5 \K. 
\end{eqnarray*}
However,  at  the low density  limit mostly  the ground  level will be
populated  and equ.   (\ref{equ:oTOpR})   will determine the ortho  to
para-\HH ratio.

Assuming that we have a constant total number density of molecular
hydrogen, $n_2$, the rate equation for para-\HH, $n_2^p$ simply
becomes,
\begin{eqnarray}
\frac{dn_2^p}{dt} = n_2^o n x k_{op},
\end{eqnarray}
where  $n_2^o$, $n$, $x$,  denote  the ortho-\HH  number  density, the
neutral   hydrogen   number   density,   and  the    ionized  fraction
respectively. The time evolution of the ionized fraction is determined
from
\begin{eqnarray}
\frac{dx}{dt} = - \k{2} n x^2
\end{eqnarray} 
to be
\begin{eqnarray}\label{equ:xoft}
x(t) = {x_0\over 1+x_0\k{2} n t}.
\end{eqnarray}
Using $n_2 = n_2^o + n_2^p$, and equ. (\ref{equ:xoft}),
we derive the simple solution for the number density of para-\HH
\begin{eqnarray}
n_2^p = n_2 \left( 1 - \frac{3}{4} {x_0}^{k_{op}/k_2} x^{-k_{op}/k_2}\right) =
        n_2 \left( 1 - \frac{3}{4} x_0^{k_{op}/k_2} \ 
          \left( n_H k_2 t + \frac{1}{x_0} \right)^{k_{op}/k_2}\right).
\end{eqnarray}
From this we clearly can see that if
\begin{itemize}
\item{k$_{op} \ll $k$_2$}, ortho-\HH to para-\HH will not change within
one recombination time.
\item{k$_{2} \ll $k$_{op}$}, ortho-\HH to para-\HH will change
dramatically within one recombination time.
\end{itemize}
Since    k$_2$    is   well  fitted    by    $  1.8    \times 10^{-10}
T^{-0.65}$cm$^3$s$^{-1}$, it convincingly shows that always the latter
case applies.   Obviously  the recombination time scale   also sets the
formation  time scale of \HH since the  electrons act as katalysts (see
also Abel 1995, and Tegmark \etal 1996). Hence when \HH gets formed by
the  gas phase  reactions (7)  through (10),  it might immediately get
converted to its para  configuration.   The subsequent cooling  of the
gas will therefore   happen  mostly in  the  $E_2$  -- $E_0$   line of
para-\HH.  For completely neutral   hydrogen gas, however, at the  low
density and temperature limit the ortho-\HH to  para-\HH ratio will be
given by $9 \exp(-170.5\K/T)$ (Mandy \& Martin 1993).  Sun \& Dalgarno
(1994) computed the rate coefficients for odd transitions of low lying
rotational levels    by  collisions  with atomic   hydrogen,   in  the
temperature range $30\K-1000\K$.  For temperatures below $300\K$ these
reactions have rate coefficients $\ll 10^{-17}\cm^3 s^{-1}$, i.e.  for
such low temperatures  in low density  gas ($n_H \sim 1 \cm^{-3}$  the
corresponding reaction time scales are $\gg 10$Gyrs.  Hence if the gas
can cool  faster than it recombines below  $300\K$ the above result is
unchanged.   However, the ortho-\HH  to para-\HH  will be different in
different environments.   Once the complete  set  of rate coefficients
for collisional excitation and  dissociation   of \HH molecules  by  H
atoms, given  by Martin \& Mandy (1995),  are extended to temperatures
below $450\K$, and extended to \HH  - \HH collisions  as well as \HH -
\Hp reactions, it will be possible  to study the ortho-\HH to para-\HH
ratio  in primordial gas  for the entire  range in  which our reaction
network is applicable.  In  summary, we would like  to stress the need
for a   fit  to the \HH  cooling   function that depends  not  only on
temperature and density, but also on the ortho-\HH to para-\HH ratio.

\clearpage

\clearpage

\clearpage

\clearpage

\clearpage

\clearpage


\begin{references}
\reference Aggarwal, K.M. 1983, \mnras, 202, 15P-20P.
\reference Aldrovandi, S.M.V.  \& Pequignot, D. 1973, \aap, 25, 137.
\reference Allison, A.C. and Dalgarno, A. 1969, Atomic Data, 1, 91. 
\reference Allison, A.C. and Dalgarno, A. 1970, 
        {\it Atomic Data},  1, 289-304.
\reference Anninos, P., Norman, M.L. 1994, \apj, 429, 434.
\reference Anninos, P, Norman, M.L. 1996, \apj, 460, 556.
\reference Anninos, P., Zhang, Y., Abel, T., and Norman, M.L. 1995, 
           {\em submitted to NewA (AZAN96).}
% \reference Anninos, W.Y., Norman, M.L. 1994, \apj, 429, 434.
% \reference Bardeen, J.M., Bond, J.R., Kaiser, N., Szalay, A.S. 1986,
           \apj 304, 15-61.
%\reference de Bernardis, P., et al. 1993, \aap. 269, 1.
\reference Bieniek, R.J. 1980, J. Phys. B, 13, 4405.
\reference Black, J.H. 1981, \mnras, 197, 553-563.
%\reference B\"{o}rner, G. 1992, {\it The Early Universe}, Springer Verlag,
%               Berlin, Heidelberg, New York.
%\reference Bond, J.R., Efstathiou, G. 1984, L45-L48.
\reference Browne, J.C. and Dalgarno, A. 1969,
         {\it J. Phys. B}, 2, 885.
%\reference Bunn, E.F., Scott, D., and White, M. 1995, \apj, 441, L9-L12.
\reference Capuzzo-Dolcetta, R., Di Fazio, A., and Palla, F. 1985,
        \aap, 145, 290-295.
%\reference Carr, B.J., Bond, J.R., and Arnett, W.D. 1984, \apj, 277, 445-469.
\reference Cen, R. 1992, \apjs, 78, 341.
\reference Corrigan, S.J.B. 1965, J. Chem. Phys., 43, 4381.
\reference Couchman, H.M.P. 1985, 
        \mnras, 214, 137-139.  
\reference Couchman, H.M.P. and Rees, M.J. 1986, 
        \mnras, 221, 53.  % Couchman-Rees star clusters
\reference Dalgarno, A. and Lepp, S. 1987, in {\it Astrochemistry}, 
        eds. Vardya, M.S. and Tarafdar, S.P. (Dordrecht:Reidel), 109-120.
\reference De Jong, T. 1972, \aap, 20, 263.
%\reference Dekel, A. and Silk, J. 1986, \apj, 303, 39-55.

\reference Dolder, K. \& Peart, B. 1985, Rep. Prog. Phys. 48, 1283-1332.
\reference Donahue, M. \& Shull, J.M. 1991, \apj, 383, 511-523.
\reference Dove, J.E. and Mandy, M.E. 1986, \apj, 311, L93-96.
\reference Dove, J.E., Mandy, M.E., Sathyamurthy, N. and Joseph, T. 1986, 
        {\it Chemical Physics Letters} Vol. 127, 1.
\reference Dove, J.E., Rusk, A.C.M., Cribb, P.H., and Martin, P.G. 1987,
         \apj, 318, 379-391.
\reference Dunn, G.H. 1968, {Phys. Rev.}, 171, p1.
\reference Ferland, G.J., Peterson, B.M., Horne, K., Welsh, W.F., 
        and Nahar, S.N. 1992, \apj, 387, 95-108.
%\reference Giroux, M.L. and Shapiro, P.R. 1994, 
%       Preprint of Univ. of Texas \#278.
%\reference Guth, A. H. {\it Inflationary universe: a possible solution
%to the horizon and flatness problem}. \pr, {23}, (1981) 347.
%\reference Guth, A., Weinberg, E. 1981, \prd, 23, 826.
%\reference Haiman, Z. 1995, Personal Communication.
\reference Haiman, Z., Thoul, A.A., \& Loeb, A. 1996, \apj, 464, 523.
\reference Hirasawa, T. 1969, {\it Progress of Theoretical Physics}, 
        Vol. 42, No. 3, 523.
%\reference Hogan, C.J., \& Rees, M.J. 1988, {Phys. Lett. B}, 205, 228.
\reference Hollenbach, D. and McKee, C.F. 1979, \apjs , 342, 555-592.
\reference Hollenbach, D. and McKee, C.F. 1989, \apj, 342, 306.
\reference Hummer, D.C., Stebbings, R.F., Fite, W.L., and Branscomb, L.M.,
        1960, {\it Phys. Rev.}, 119, 668.
\reference Hutchins, J.B.,
        1976, \apj, 205, 103-121.
\reference Janev, R.K., Langer, W.D., Evans, Jr. K. and 
        Post, Jr. D.E. 1987, 
        {\it Elementary Processes in Hydrogen-Helium Plasmas} 
        (Springer-Verlag).
\reference Jura, M. 1974, \apj, 191, 378.
\reference Kang, H. and Shapiro, P.R. 1992, \apj, 386, 432.
\reference Karpas, Z., Anicich, V., \& Huntress, W.T.Jr. 1979,
        J. Chem. Phys. 70, 2877.
\reference Kashlinsky, A. and Rees, M.J. 1983, \mnras, 205, p955-971.
%\reference Kolb, E.W. \& Turner, M. 1990, {\it The Early Universe},
%       Addison-Wesley, Redwood City/California. 
\reference Launay, J.M., Le Dourneuf, M., \& Zeippen, C.J. 1991, \aap,
252, 842-852.
\reference Lenzuni, P., Chernoff, D.F. and Salpeter,E.E. 1991,  
        \apjsupp, 76, 759.
\reference Lepp, S., and Shull, J.M. 1983, \apj, 270, 578-582.
%\reference Leslie, J. 1989, {\it Universes}, Routledge, London, New York.
%\reference Linde, A. 1982, Phys. Let., 108B, 389.
%\reference Linde, A. 1990, {\it Particle Physics and Inflationary
%       Cosmology}, Harvard Acad. Publ..
%\reference Mac Low, M.-M. 1995, Personal Communication.
\reference Mac Low, M.-M. and Shull, J.M. 1986, \apj, 302, 585.
\reference Mandy, M.E., \& Martin, P.G. 1993, \apjs, 86, 199.
\reference Martin, P.G., Schwarz, D.H., \& Mandy, M.E. 1996, \apj, 
  461, 265.
\reference Mitchell, G.F. and Deveau, T.J. 1983, \apj, 266, 646-661.
\reference Moseley, J., Olson, R.E., and Peterson, J.R. 1975, {\it
        Case Studies in Atomic Physics}, vol.5 (Amsterdam:
        North-Holland), 1.
\reference Nakashima, K., Takagi, H., and Nakamura, H. 1987,
        J. Chem. Phys., Vol. 86, 726-737.
%\reference Navarro, J.F., Frenk, C.S., White, S.D.M. 1995, MPA
%           Preprint 884, (submitted to ApJ).
\reference Osterbrock, D.E. 1974, {\it Astrophysics of Gaseous Nebulae},
        Freeman and Co., San Francisco.
\reference Osterbrock, D.E. 1989, {\it Astrophysics of Gaseous Nebulae 
        and Active Galactic Nuclei.},
        University Science Books, Mill Valley/California.
\reference Padmanabhan, T. 1993, {\it Structure formation in the
        universe}, Cambridge University Press.
%\reference Palla, F., Salpeter, E.E., \& Stahler, S.W. 1983, \apj,
%       271, 632-641.
\reference Peart, B., Bennett, M.A., and Dolder, K. 1985, 
        Phys. B, 18, L439.
%\reference Peebles, P.J.E. 1971, {\it The Large Scale Structure of
%      the Universe},           Princeton University Press, Princeton.
%\reference Peebles, P.J.E. 1982, \apj, 263, L1-L5.
%\reference Peebles, P.J.E. 1993, {\it Physical Cosmology}, 
%       Princeton University Press, Princeton.
\reference Peebles, P.J.E. and Dicke, R.H. 1968, \apj, 154, 891.
\reference Poulaert, G., Brouillard, F., Claeys, W., McGowan, J.W., and 
  Van Wassenhove, G. 1978, 
        J. Phys. B: Atom. Molec. Phys., Vol. 11, L671-673.
\reference Prasad, S.S., \& Huntress, W.T.,Jr. 1980, \apjs, 43, 1.
\reference Puy, D., Alecian, G., Le Bourlot, J., Leorat, J., \& Pineau des
                   Forets, G. 1993, \aap, 267, 337-346.
\reference Ramaker, D.E., and Peek, J.M. 1976, \pra, 13, 58.
\reference Rawlings, J.M.C., Drew, J.E. and Barlow,, M.J. 1993, 
        \mnras, 265, 968-982.
\reference Rees, M.J. 1986, \mnras, 222, 27P-32P.
%\reference Rees, M.J., Ostriker, J.P. 1977, \mnras, 179, 541-559
\reference Rosenzweig, P., Parravano, A., Ib\'{a}\~{n}ez,M.H. 
        and Izotov, Yu.I. 1994, \apj, 432, 485.
\reference Saslaw, W.C. and Zipoy, D. 1967, Nature, 216, 976-978.
%\reference  Sato, K. 1981, \mnras, 195, 487.
\reference Schneider, I.F., Dulieu, O., Giusti-Suzor, A., \& Roueff, E.,
        1994, \apj, 424, 983-987.
%\reference Seidel, E., Suen, W.M. 1994, \prl, 72, 2516-2519.
\reference Senekowitsch, J., Rosmus P., Domcke, W. \& Werner,
H.J. 1984, Chem. Phys. Lett., 111, 211.
\reference Shapiro, P.R. and Kang, H. 1987, \apj, 318, 32.
\reference Shull, J.M. 1978, \apj, 219, 877-885.
\reference Shull, J.M. and McKee, C.F. 1979, \apj, 227, 131-149.
\reference Stahler, S.W. 1986, \pasp, 98, 1081.
\reference Stancil, P.C. 1994, \apj, 430, 360-370.
\reference Stancil, P.C., Lepp, S., Dalgarno, A. 1996, \apj, 458, 401-406.
\reference Stecher, T.P.,\& Williams, D.A. 1967, \apj, 149, L29.
%\reference Stone, J.M., \& Norman, M.L. 1992, \apjs, 80, 753.
\reference Sun, Y. \& Dalgarno, A. 1994, \apj, 427, 1053-1056.
\reference Szucs, S., Karemara, M., Terao, M. 1983, {\it Proc. 13th
        Int. Conf.\ on Physics of Electionic and Atomic Collisions}, Berlin,
        ed. Eichler et al., (Amsterdam: North-Holland) Abstracts, 482.
\reference Szucs, S., Karemara, M., Terao, M., Brouillard, F.J., 
        1984, Phys. B, 17, 1631.
\reference \href{http://www.mpa-garching.mpg.de/~max/minmass.html}{
Tegmark, M., Silk, J., Rees, M. J., Blanchard, A., Abel, T., 
Palla, P. 1996, to appear in \apj, 473.}
\reference Tielens, A.G.G.M. and Hollenbach,D. 1985, \apj, 291, 722.
\reference \href{http://www.pa.uky.edu/~verner/atom.html}{Verner Dima,
1996}
\reference Vietri, M., \& Pesce, E. 1995, \apj, 442, 618-627.
\reference Wadehra, J.M. and Bardsley, J.N. 1978, \prl, 41, 1795-1798.
\reference Wishart, 1979, \mnras, 187, 59P-60P.
%\reference Zel'dovich, Ya.B. 1980, \mnras, 192, 663-667.
\reference Zhang, Y. \& Anninos, P., \& Norman, M.L. 1995, \apjl, 453, 57Z.
\reference Zhang, Y., Anninos, P., Abel, T., \& Norman, M.L. 1996, 
           {\em submitted to NewA.}
\end{references}
\end{document}